\documentclass[11pt,a4paper]{article}

\usepackage{amsmath,amssymb,tikz}
\usepackage{graphicx}
 \allowdisplaybreaks
\def\be{\begin{equation}}
\def\ee{\end{equation}}
\def\ba#1\ea{\begin{align}#1\end{align}}
\def\bg#1\eg{\begin{gather}#1\end{gather}}
\def\bm#1\em{\begin{multline}#1\end{multline}}
\def\bmd#1\emd{\begin{multlined}#1\end{multlined}}

\def\({\left(}
\def\){\right)}
\def\[{\left[}
\def\]{\right]}

\def \be {\begin{equation}}
\def \ee {\end{equation}}
\def \ba {\begin{array}}
\def \ea {\end{array}}
\def \bea{\begin{eqnarray}}
\def \eea{\end{eqnarray}}

\def\bea{\begin{eqnarray}}
\def\eea{\end{eqnarray}}

\newcommand{\bit}{\begin{itemize}}  \newcommand{\eit}{\end{itemize}}
\newcommand{\ben}{\begin{enumerate}}  \newcommand{\een}{\end{enumerate}}

\long\def\symbolfootnote[#1]#2{\begingroup%
\def\thefootnote{\fnsymbol{footnote}}\footnote[#1]{#2}\endgroup}

\newcommand{\sysu}{{\it School of Physics and Astronomy, Sun Yat-Sen University, 2 Daxue Road, Zhuhai 519082, China}}

\begin{document}
\thispagestyle{empty}
\begin{center}

~\vspace{20pt}

{\Large\bf Holographic Entanglement Entropy for Brane-World Higher Derivative Gravity}

\vspace{25pt}

Rong-Xin Miao ${}$\symbolfootnote[1]{Email:~\sf
  miaorx@mail.sysu.edu.cn}, Zi-Bing Xie ${}$\symbolfootnote[2]{Email:~\sf
  xiezb@mail2.sysu.edu.cn}

\vspace{10pt}${}$\sysu

\vspace{2cm}

\begin{abstract}
Due to the splitting problem, it is difficult to derive the holographic entanglement entropy for general higher derivative gravity. Inspired by double holography and renormalized entanglement entropy, we develop a method to derive the generalized gravitational entropy for the brane-world higher derivative (BWHD) gravity. Remarkably, this approach is independent of the splitting problem. The so-called BWHD gravity is an effective theory on the brane, given by the counter terms of holographic renormalization. Interestingly, all solutions to Einstein gravity are also solutions to BWHD gravity. We first verify our approach can derive the correct results for curvature-squared gravity and then derive the holographic entanglement entropy for cubic BWHD gravity, which is the main result of this paper. We also derive the entropy of quartic BWHD gravity in flat space with constant extrinsic curvatures and perform several tests on our results. Finally, we briefly comment on our results. 
\end{abstract}

\end{center}

\newpage
\setcounter{footnote}{0}
\setcounter{page}{1}

\tableofcontents

\section{Introduction}

The developments of AdS/CFT correspondence \cite{Maldacena:1997re, Witten:1998qj} provide several modern insights into the origin of gravity. First, it proposes gravity is a holographic theory \cite{tHooft:1993dmi, Susskind:1994vu}, which is dual to a conformal field theory (CFT) without gravity on the boundary. Second, it suggests a deep relation between geometry and quantum entanglement according to the Ryu-Takayanagi formula \cite{Ryu:2006bv}. The Ryu-Takayanagi formula, also called holographic entanglement entropy (HEE) and generalized gravitational entropy, is a generalization of the Bekenstein-Hawking entropy for black holes, which reveals the quantum information origin of gravity. The recent breakthrough in the black hole information problem \cite{Penington:2019npb, Almheiri:2019psf, Almheiri:2019hni} also benefits from the developments of HEE, in particular, the concept of quantum extremal surfaces \cite{Faulkner:2013ana,Engelhardt:2014gca}. This paper studies the HEE for higher derivative (HD) gravity. String theory predicts the effective theory of quantum gravity is an HD gravity \cite{Gross:1986mw, Gross:1986iv, Fradkin:1984pq}. Besides, HD gravity is renormalizable generally \cite{Stelle:1976gc, Buchbinder:1992rb} and can describe more general gravity duals of strongly coupled CFTs \cite{Buchel:2009sk, Gregory:2009fj}. Thus, there is an excellent motivation to investigate the HEE of HD gravity. 

There are many methods to study the HEE of HD gravity. We discuss only two typical methods. The first is the squashed-cone method developed to prove the RT formula \cite{Lewkowycz:2013nqa}. By applying it, Dong and Camps independently obtain the HEE for higher curvature gravity \cite{Dong:2013qoa, Camps:2013zua}, which contributes anomalous corrections involving extrinsic curvatures to the famous Wald entropy \cite{Wald:1993nt}. Later, it is found, due to the splitting problem \cite{Miao:2014nxa, Miao:2015iba}, there is arbitrariness in the entropy formulas of \cite{Dong:2013qoa, Camps:2013zua}. In principle, the splitting problem and, thus, the entropy arbitrariness can be fixed by solving HD gravity equations of motion. However, it is a non-trivial problem for general HD gravity. See \cite{Miao:2014nxa, Miao:2015iba, Dong:2017xht, Camps:2016gfs} for more discussions of the splitting problem. The second method is the second-law approach \cite{Wall:2015raa}. Wall observes that the second law of black hole entropy can fix the squared terms of extrinsic curvatures in the entropy of dynamical black holes \cite{Wall:2015raa}. Wall's approach has recently been generalized to higher orders of extrinsic curvatures in effective field theory (EFT) \cite{Davies:2022xdq, Davies:2023qaa}. This method generally depends on EFT's validity and requires small HD couplings. Due to the above difficulties and limitations, so far, we only know the non-perturbative covariant formulas of HEE for Einstein gravity \cite{Ryu:2006bv, Lewkowycz:2013nqa}, curvature-squared gravity \cite{Dong:2013qoa, Camps:2013zua}, Gauss-Bonnet/Lovelock gravity \cite{Hung:2011xb,deBoer:2011wk}, gravitational Chern-Simons terms \cite{Guo:2015uqa, Azeyanagi:2015uoa} and some kinds of cubic gravity in six dimensions \cite{Miao:2015iba}. See also \cite{Bueno:2020uxs, Caceres:2020jrf} for the perturbative HEE of HD gravity and the gravitational entropy from Hayward corner term \cite{Kastikainen:2023omj, Kastikainen:2023yyk}. 

Inspired by double holography \cite{Chen:2020uac} and renormalized holographic entanglement entropy (RHEE) \cite{Taylor:2016aoi}, we develop a new approach to derive the HEE for brane-world higher derivative (BWHD) gravity. The so-called BWHD gravities are the counterterms of holographic renormalization \cite{Balasubramanian,deHaro:2000vlm}, which are identities as the effective action of ``quantum gravity" on the branes in double holography \cite{Chen:2020uac}. Interestingly, the BWHD gravity can construct a new kind of massive gravity in three dimensions \cite{Bueno:2023dpl}. Our method can reproduce the HEE for curvature-squared gravity \cite{Dong:2013qoa, Camps:2013zua}. Furthermore, it drives a new non-perturbative covariant entropy formula for the cubic BWHD gravity in general dimensions. Finally, we briefly discuss the HEE for quartic BWHD gravity in flat space. We make several tests of our results. Remarkably, our approach is independent of the splitting problem. Note that BWHD gravities also suffer the splitting problem: different splittings of squashed cones yield different entropy. Our approach bypasses this problem by the self-consistency of double holography or the finiteness of renormalized entanglement entropy. However, it can only deal with the BWHD gravity. We remark that BWHD gravity is a fascinating HD theory of gravity. First of all, all solutions to Einstein gravity are also solutions to BWHD gravity. Thus, it can be regarded as an HD generalization of Einstein gravity. Besides, as an effective theory of ``quantum gravity" on the branes, it can help us to explore novel quantum effects of gravity \cite{Emparan:2020znc, Panella:2024sor} and study the black hole information paradox in double holography \cite{Chen:2020uac}.

The paper is organized as follows. Section 2 formulates our approach based on double holography and RHEE. To test our approach, we reproduce the HEE of curvature-squared gravity. In section 3, we derive the HEE of cubic BWHD gravity by applying our method. This is the main result of this paper. Section 4 tests the cubic entropy formula by comparing it with Wald entropy, HEE without/with extrinsic curvatures, and HEE under field redefinition. Section 5 studies the HEE of quartic  BWHD gravity in flat space with constant extrinsic curvatures. Finally, we conclude with some discussions in section 6. 

Notations: we abridge `entanglement entropy', `holographic entanglement entropy', `renormalized holographic entanglement entropy', and `brane-world higher derivative' as EE, HEE, RHEE, and BWHD, respectively. The other notations for bulk $N$, AdS boundary $M=\partial N$, bulk entangling surface $\Gamma$, and boundary entangling surface $\gamma=\partial \Gamma$ are listed in Table. \ref{sect 2: notation123}. 

\begin{table}[!htbp]\label{sect 2: notation123}
	\centering
	\renewcommand{\arraystretch}{1.6}
	\begin{tabular}{|c|c|c|c|c|}
		\hline
		& bulk $N$ & AdS bdy $M$ & bulk entangling surface $\Gamma$ & $\gamma=\partial\Gamma$ \\
		\hline
		dimension & d+1 & d & d-1 & d-2 \\
		\hline
		coordinate & $X^{\mu}=(\rho,x^i)$ & $x^i$ & $Y^{\hat{\alpha}}=(\tau,y^{\hat{i}})$ & $y^{\hat{i}}$\\
		\hline
		metric & $\hat{G}_{\mu\nu}$ & $\tilde{g}_{ij}=\frac{1}{\rho}g_{ij}=\frac{1}{\rho}\overset{(0)}{g}_{ij}+\dots$ & $\hat{h}_{\hat{\alpha}\hat{\beta}}$ & $\hat{h}_{\hat{i}\hat{j}}$\\
		\hline
		metric det& $\hat{G}$ & $\tilde{g}$ & $H$ & $h$\\
		\hline
		curvature & $\hat{R}$ & $\tilde{R}$ & $\hat{r}$ & $r$\\
		\hline
	\end{tabular}
	\caption{Notations}
\end{table}

\section{Approach to derive HEE}

Inspired by RHEE and double holography, we develop a new method to derive the HEE for the brane-world higher derivative (BWHD) gravity in this section. For simplicity, we take the curvature-square gravity as an example to illustrate the main ideas. We verify that our method can produce the correct HEE for curvature-square gravity. 

The naked gravitational action in AdS/CFT reads
 \begin{eqnarray}\label{sect 2: bulk action}
I_0=\int_Nd^{d+1}X\sqrt{|\hat{G}|}\Big(\hat{R}+d(d-1)\Big)+2 \int_M d^{d}x\sqrt{|\tilde{g}|} K
 \end{eqnarray}
 where $\hat{G}_{\mu\nu}$ and $\hat{R}$ are the metric and Ricci scalar in bulk $N$, $\tilde{g}_{ij}$ and $K$ are the induced metric and extrinsic curvatures on the AdS boundary $M$. We have set Newton's constant $16\pi G_N=1$ and the AdS radius $l=1$ for simplicity. To have a finite gravitational action, one should perform the holographic renormalization by adding suitable counterterms on the AdS boundary $M$ \cite{Balasubramanian,deHaro:2000vlm}
\begin{align} \label{sect 2: counterterms}
	I_c=\int_Md^{d}x\sqrt{|\tilde{g}|} \bigg[-2(d-1)-\frac{1}{d-2}\tilde{R}-\frac{\mathcal{L}_2}{(d-4)(d-2)^2}
	+\frac{2\mathcal{L}_3}{(d-6)(d-4)(d-2)^3}+\dots\bigg],
\end{align}
where $\mathcal{L}_2$ and $\mathcal{L}_3$ are given by \cite{Kraus:1999di}
\begin{equation} \label{sect 2: L2}
	\mathcal{L}_2=\tilde{R}{}^{ij}\tilde{R}_{ij}-\frac{d}{4(d-1)}\tilde{R}^2,
\end{equation}
and
\begin{equation} \label{sect 2: L3}
	\mathcal{L}_3=\frac{3d+2}{4(d-1)}\tilde{R}\tilde{R}^{ij}\tilde{R}_{ij}-\frac{d(d+2)}{16(d-1)^2}\tilde{R}^3-2\tilde{R}^{ij}\tilde{R}_{ikjl}\tilde{R}^{kl}+\frac{d}{4(d-1)}\tilde{R}\Box \tilde{R}-\tilde{R}^{ij}\Box \tilde{R}_{ij}.
\end{equation}
In this way, we get a finite renormalized gravitational action
\begin{equation} \label{sect 2: Iren ICFT}
I_{\text{ren}}=I_{\text{CFT}}=I_0+I_c,
\end{equation}
which is equivalent to the effective action of dual CFTs, i.e., $I_{\text{CFT}}=-\log Z_{\text{CFT}}$, with $Z_{\text{CFT}}$ the renormalized CFT partition function. Since $I_{\text{ren}}=I_0+I_c$ is finite, all the physical quantities of CFTs derived from it are finite too. The HEE derived from $I_{\text{ren}}$ is called RHEE \cite{Taylor:2016aoi}. See also the RHEE derived from Kounterterms \cite{Anastasiou:2017xjr}. We turn around the logic of RHEE \cite{Taylor:2016aoi} in this paper. Instead of getting a finite RHEE, we make use of the finiteness of RHEE to derive the generalized gravitational entropy of BWHD gravity on the AdS boundary. The main ideas are as follows. The RHEE obtained from $I_{\text{ren}}=I_0+I_c$ (\ref{sect 2: bulk action},\ref{sect 2: counterterms}) takes the form
\begin{align} \label{sect 2: RHEE}
	S_{\text{RHEE}}=&4\pi \int_\Gamma d^{d-1}Y\sqrt{H} \notag\\
	&+\int_{\gamma=\partial \Gamma} d^{d-2}y\sqrt{h}\bigg[-\frac{4\pi}{d-2}-\frac{S(\mathcal{L}_2)}{(d-4)(d-2)^2}+\frac{2S(\mathcal{L}_3)}{(d-6)(d-4)(d-2)^3}+...\bigg],
\end{align}
where $S(\mathcal{L}_2)$ and $S(\mathcal{L}_3)$ denote the generalized gravitational entropy for $\mathcal{L}_2$ (\ref{sect 2: L2}) and $\mathcal{L}_3$ (\ref{sect 2: L3}), $\Gamma$ and $\gamma=\partial \Gamma$ label the entangling surfaces in bulk $N$ and on the AdS boundary $M$, respectively. See Table. 1 for various notations. We expand $S_{\text{RHEE}}$ in terms of the cut-off $\rho=z^2=\epsilon^2$ near the AdS boundary $\rho=z^2\to 0$ and require the divergent terms of $S_{\text{RHEE}}$ vanish. In this way, we can derive the non-perturbative expressions of $S(\mathcal{L}_2), S(\mathcal{L}_3)$ and so on order by order.

Let us reexplain the above approach in the language of double holography. The double holography (or AdS/BCFT) states that Einstein gravity coupled with a brane in bulk is dual to ``quantum gravity" on the brane and is dual to a conformal field theory with a boundary (BCFT) on the AdS boundary. The bulk action in double holography or AdS/BCFT reads 
 \begin{eqnarray}\label{sect 2: action in AdS/BCFT}
I_{\text{bulk}}=I_0-2 \int_{M} d^{d}x\sqrt{|\tilde{g}|} T,
 \end{eqnarray}
where $I_0$ is (\ref{sect 2: bulk action}) and $T$ is the brane tension. In double holography, the brane $M$ is located at finite place $\rho=z^2=\epsilon^2>0$ instead of the AdS boundary $\rho=z^2=0$. As a result, $I_{\text{bulk}}$ is finite unless the brane is located at infinity with $\epsilon=0$. By using $I_0=I_{\text{CFT}}-I_c$ (\ref{sect 2: Iren ICFT}), we rewrite (\ref{sect 2: action in AdS/BCFT}) as
 \begin{eqnarray}\label{sect 2: action in AdS/BCFT 2}
I_{\text{brane}}=I_{\text{CFT}}-I_c-2 \int_{M} d^{d}x\sqrt{|\tilde{g}|} T.
 \end{eqnarray}
According to \cite{Chen:2020uac}, (\ref{sect 2: action in AdS/BCFT 2}) is the effective action of ``quantum gravity" on the brane with $(-I_c)$ the so-called BWHD gravity. Although HD gravity suffers the ghost problem generally, the total action (\ref{sect 2: action in AdS/BCFT 2}) is ghost-free since it is derived from Einstein gravity \cite{Hu:2022lxl}. The CFT action $I_{\text{CFT}}$ plays a vital role in removing the ghost of BWHD gravity \cite{Hu:2022lxl}.

Note that (\ref{sect 2: action in AdS/BCFT}) includes a bulk action $I_0$, while the (\ref{sect 2: action in AdS/BCFT 2}) contains only the action on the brane. They correspond to the bulk and brane perspectives of double holography, respectively. From the action (\ref{sect 2: action in AdS/BCFT}) in the bulk perspective, we derive the HEE
 \begin{eqnarray}\label{sect 2: HEE bulk}
S_{\text{bulk}}=4\pi \int_\Gamma d^{d-1}Y\sqrt{H},
 \end{eqnarray}
 which is the first line of (\ref{sect 2: RHEE}). 
 Note that the tension term does not contribute to HEE since it contains no curvatures. Similarly, from the action (\ref{sect 2: action in AdS/BCFT 2}) in the brane perspective, we obtain the HEE
  \begin{eqnarray}\label{sect 2: HEE bulk}
S_{\text{brane}}=\int_{\gamma=\partial \Gamma} d^{d-2}y\sqrt{h}\bigg[\frac{4\pi}{d-2}+\frac{S(\mathcal{L}_2)}{(d-4)(d-2)^2}-\frac{2S(\mathcal{L}_3)}{(d-6)(d-4)(d-2)^3}+...\bigg]+ O(\epsilon^0)
 \end{eqnarray}
which is the negative of the second line of (\ref{sect 2: RHEE}). Note that $I_{\text{CFT}}$ contributes to only the finite term $O(\epsilon^0)$ of HEE. According to double holography, we have $S_{\text{bulk}}=S_{\text{brane}}$. Now we take the limit $\epsilon\to 0$ (brane approaches AdS boundary) and require the divergent terms of $S_{\text{bulk}}$ and $S_{\text{brane}}$ match. Equivalently, we require the RHEE $(S_{\text{RHEE}}=S_{\text{bulk}}-S_{\text{brane}})$ is finite for $\epsilon \to 0$.

Above, we have explained our method of deriving HEE in RHEE and double holography viewpoints, respectively. Below, we take $S(\mathcal{L}_2)$ as an example to illustrate the technical details of our method. First, we give a quick review of the Fefferman-Graham expansion and the Schwimmer-Theisen approach \cite{Schwimmer:2008yh}, which are helpful for the derivations of HEE.

\subsection{Fefferman-Graham expansion}

The metric of an asymptotically AdS spacetime can be expanded in the Fefferman-Graham gauge
\begin{equation}\label{sect 2:bulk metric}
	ds_N^2=\hat{G}_{\mu\nu}dX^{\mu}dX^{\nu}=\frac{1}{4\rho^2}d\rho^2+\tilde{g}_{ij}dx^idx^j=\frac{1}{4\rho^2}d\rho^2+\frac{1}{\rho}g_{ij}dx^idx^j,
\end{equation}
where $\rho=z^2\to 0$ denote the AdS boundary and 
\begin{equation}\label{sect 2:FG gauge}
	g_{ij}=\overset{(0)}{g}_{ij}+\rho\overset{(1)}{g}_{ij}+...+\rho^{\frac{d}{2}}(\overset{(\frac{d}{2})}{g}_{ij}+\overset{(\frac{d}{2})}{\gamma}_{ij} \log\rho)+....
\end{equation}
Remarkably, $\overset{(n)}{g}_{ij}$ can be universally determined by the PBH transformation \cite{Imbimbo:1999bj}, which gives
\begin{equation}\label{sect 2:g1ij}
	\overset{(1)}{g}_{ij}=-\frac{1}{d-2}(\overset{(0)}{R}_{ij}-\frac{\overset{(0)}{R}}{2(d-1)}\overset{(0)}{g}_{ij}).
\end{equation}
and 
\begin{align}\label{sect 2:g2ij}
	\overset{(2)}{g}_{ij}=&k_1C_{mnkl}C^{mnkl}\overset{(0)}{g}_{ij}+k_2C_{iklm}C_j{}^{klm}\notag\\
	&+\frac{1}{d-4}\bigg[\frac{1}{8(d-1)}\nabla_i\nabla_j\overset{(0)}{R}-\frac{1}{4(d-2)}\Box \overset{(0)}{R}_{ij}+\frac{1}{8(d-1)(d-2)}\Box \overset{(0)}{R}\overset{(0)}{g}_{ij}\notag\\
	&-\frac{1}{2(d-2)}\overset{(0)}{R}{}^{kl}\overset{(0)}{R}_{ikjl}+\frac{d-4}{2(d-2)^2}\overset{(0)}{R}{}_i{}^k\overset{(0)}{R}_{jk}+\frac{1}{(d-1)(d-2)^2}\overset{(0)}{R}\overset{(0)}{R}_{ij}\notag\\
	&+\frac{1}{4(d-2)^2}\overset{(0)}{R}{}^{kl}\overset{(0)}{R}_{kl}\overset{(0)}{g}_{ij}-\frac{3d}{16(d-1)^2(d-2)^2}\overset{(0)}{R}{}^2\overset{(0)}{g}_{ij}\bigg],
\end{align}
where $k_1=k_2=0$ for our case of Einstein gravity. By dimensional analysis, we know that $S(\mathcal{L}_2)$ is of order $\overset{(1)}{g}_{ij}$ and $S(\mathcal{L}_3)$ is of order $\overset{(2)}{g}_{ij}$. 

\subsection{Schwimmer-Theisen approach}

Label the embedding function of (d-1)-dimensional bulk entangling surface $\Gamma$ by  $X^{\mu}=X^{\mu}(Y^{\hat{\alpha}})$, where $X^\mu=\{x^i,\rho\}$ are bulk coordinates and $Y^{\hat{\alpha}} =\{y^{\hat{i}},\tau\}$ are  coordinates on $\Gamma$.  The induced metric on the bulk entangling surface becomes
\begin{equation}\label{sect 2: induced metric on Gamma}
\hat{h}_{\hat{\alpha}\hat{\beta}}=\partial_{\hat{\alpha}} X^\mu\partial_{\hat{\beta}} X^\nu \hat{G}_{\mu\nu}.
\end{equation}
We choose the Schwimmer-Theisen gauge 
\begin{equation}\label{sect 2:chosen gauge}
	\tau=\rho,\quad \hat{h}_{\tau \hat{i}}=0,
\end{equation}
and expand the embedding functions in series of $\tau$
\begin{equation}\label{sect 2:embedding functions}
	X^{i}(\tau,y^{\hat{i}})=\overset{(0)}{X}{}^{i}(y^{\hat{i}})+\overset{(1)}{X}{}^{i}(y^{\hat{i}})\tau+\overset{(2)}{X}{}^{i}(y^{\hat{i}})\tau^2+...
\end{equation}
Similar to the PBH transformation for $\overset{(n)}{g}_{ij}$ \cite{Imbimbo:1999bj}, the diffeomorphism preserving Fefferman-Graham gauge (\ref{sect 2:FG gauge}) and Schwimmer-Theisen gauge (\ref{sect 2:chosen gauge}) determines uniquely the transformation rules of $\overset{(n)}{X}{}^{i}(y^{\hat{i}})$ \cite{Schwimmer:2008yh}. From these transformation rules, we can identity $\overset{(1)}{X}{}^{i}(y^{\hat{i}})$ with the extrinsic curvature of the entangling surface $\gamma=\partial \Gamma$ on the AdS boundary
\begin{equation}\label{sect 2:X(1)}
	\overset{(1)}{X}{}^{i}(y^{\hat{i}})=\frac{1}{2(d-2)}\overset{(0)}{k}{}^i(y^{\hat{j}}),
\end{equation}
where the extrinsic curvatures are defined by
\begin{equation}\label{sect 2:extrinsic curvature}
\overset{(0)}{k}{}^i= \overset{(0)}{h}{}^{\hat{i}\hat{j}}\overset{(0)}{k}{}^i_{\hat{i}\hat{j}},\ \ \ \ \overset{(0)}{k}{}^i_{\hat{i}\hat{j}}=\partial_{\hat{i}}\partial_{\hat{j}}\overset{(0)}{X}{}^i-\overset{(0)}{\gamma}{}^{\hat{l}}_{\hat{i}\hat{j}}\partial_{\hat{l}}\overset{(0)}{X}{}^i+\overset{(0)}{\Gamma}{}^i_{jk}\partial_{\hat{i}}\overset{(0)}{X}{}^j\partial_{\hat{j}}\overset{(0)}{X}{}^k,
\end{equation}
with $\overset{(0)}{\Gamma}{}^i_{jk}$ and $\overset{(0)}{\gamma}{}^{\hat{l}}_{\hat{i}\hat{j}}$  the Levi-Civita connections with respect to $\overset{(0)}{g}_{ij}$ and $\overset{(0)}{h}_{\hat{i}\hat{j}}=\partial_{\hat{i}} \overset{(0)}{X}{}^i\partial_{\hat{j}} \overset{(0)}{X}{}^j\overset{(0)}{g}_{ij}
$ respectively.

From (\ref{sect 2:bulk metric}, \ref{sect 2:embedding functions}, \ref{sect 2:X(1)}), we derive the induced metric (\ref{sect 2: induced metric on Gamma}) on $\Gamma$
\begin{align}\label{sect 2:htautau hij}
	\hat{h}_{\tau\tau}=\frac{1}{4\tau^2}\bigg[1+\tau \overset{(1)}{h}_{\tau\tau}+O(\tau^2)\bigg],\ \ \ \hat{h}_{\hat{i}\hat{j}}	=\frac{1}{\tau}\bigg[\overset{(0)}{h}_{\hat{i}\hat{j}}+\tau \overset{(1)}{h}_{\hat{i}\hat{j}}+O(\tau^2)\bigg],
\end{align}
where 
\begin{equation}\label{sect 2:h1ij}
\overset{(1)}{h}_{\tau\tau}=\frac{1}{(d-2)^2}\overset{(0)}{k}{}^i\overset{(0)}{k}{}^j\overset{(0)}{g}_{ij}
,\  \ \ \overset{(1)}{h}_{\hat{i}\hat{j}}=\overset{(1)}{g}_{\hat{i}\hat{j}}-\frac{1}{d-2}\overset{(0)}{k}{}^i\overset{(0)}{k}{}^j_{\hat{i}\hat{j}}\overset{(0)}{g}_{ij}.
\end{equation}
The above two equations yield 
\begin{align}\label{sect 2:H}
	\sqrt{H}&=\sqrt{ |\hat{h}_{\hat{\alpha}\hat{\beta}}|}=\frac{1}{2\tau^{\frac{d}{2}}}\sqrt{\overset{(0)}{h}}\bigg[1+\frac{\tau}{2}(\overset{(1)}{h}_{\tau\tau}+\overset{(1)}{h}_{\hat{i}}{}^{\hat{i}})+O(\tau^2)\bigg],\notag\\
	&=\frac{1}{2\tau^{\frac{d}{2}}}\sqrt{\overset{(0)}{h}}\bigg[1+\frac{\tau}{2}(\overset{(1)}{g}{}_{\hat{i}}{}^{\hat{i}}-\frac{d-3}{(d-2)^2}\overset{(0)}{k}{}^i \overset{(0)}{k}{}_i)+O(\tau^2)\bigg]
\end{align}
and
\begin{align}\label{sect 2:h}
	\sqrt{h}&=\sqrt{ |\hat{h}_{\hat{i}\hat{j}}|}=\frac{1}{\tau^{\frac{d}{2}-1}}\sqrt{\overset{(0)}{h}}\bigg[1+\frac{\tau}{2}\overset{(1)}{h}_{\hat{i}}{}^{\hat{i}}+O(\tau^2)\bigg]\notag\\
	&=\frac{1}{\tau^{\frac{d}{2}-1}}\sqrt{\overset{(0)}{h}}\bigg[1+\frac{\tau}{2}(\overset{(1)}{g}{}_{\hat{i}}{}^{\hat{i}}-\frac{1}{(d-2)}\overset{(0)}{k}{}^i \overset{(0)}{k}{}_i)+O(\tau^2)\bigg].
\end{align}

\subsection{HEE for squared HD gravity}

Now we are ready to derive $S(\mathcal{L}_2)$ by applying our RHEE method.  Recall that $\mathcal{L}_2$  (\ref{sect 2: L2}) denotes the squared curvatures constructed from $\tilde{g}_{ij}=\overset{(0)}{g}_{ij}/\rho+O(1)$. As a result, we have $S(\mathcal{L}_2) \sim \tilde{R}\sim \rho \overset{(0)}{R}+O(\rho^2)$, which yields
\begin{align}\label{sect 2: SL2 relation}
S\left(\mathcal{L}_2\right)=\tau  \overset{(0)}{S}(\mathcal{L}_2)+O(\tau^2),
\end{align}
where $ \overset{(0)}{S}(\mathcal{L}_2)=S(\mathcal{L}_2(\overset{(0)}{g}))$ and $\rho=\tau$.  Substituting (\ref{sect 2:H},\ref{sect 2:h}) and (\ref{sect 2: SL2 relation}) into the RHEE (\ref{sect 2: RHEE}) and focusing on the UV divergent terms, we derive
\begin{align} \label{sect 2: derive SL2}
	S_{\text{RHEE}}&=4\pi \int_{\epsilon^2}d\tau \int_{\gamma=\partial \Gamma}d^{d-2}y\sqrt{H} +\int_{\gamma (\tau=\epsilon^2)} d^{d-2}y\sqrt{h}\bigg[-\frac{4\pi}{d-2}-\frac{S(\mathcal{L}_2)}{(d-4)(d-2)^2}+...\bigg]\notag\\
&=\int_{\gamma (\tau=\epsilon^2)} d^{d-2}y\sqrt{\overset{(0)}{h}} \frac{1}{\epsilon^{d-4}} \Big[ \frac{4 \pi  (d-2) \overset{(1)}{g}{}^{\hat{i}}{}_{\hat{i}}-2 \pi  \overset{(0)}{k}{}_i \overset{(0)}{k}{}^i- \overset{(0)}{S}(\mathcal{L}_2)}{(d-4) (d-2)^2}+O(\epsilon^2)\Big],
\end{align}
where we have choose the UV cutoff $\tau=\epsilon^2$. Recall that the RHEE (\ref{sect 2: derive SL2}) should be finite by definition. Thus, we obtain for sufficiently large $d$ ($d> 4$) 
\begin{align} \label{sect 2: get SL2}
\overset{(0)}{S}(\mathcal{L}_2)=4 \pi  (d-2) \overset{(1)}{g}{}^{\hat{i}}{}_{\hat{i}}-2 \pi  \overset{(0)}{k}{}_i \overset{(0)}{k}{}^i=2 \pi  \left(\frac{d-2}{d-1}\overset{(0)}{R} -\overset{(0)}{k}{}^i\overset{(0)}{k}{}_i-2 \overset{(0)}{R}{}^{\hat{i}}{}_{\hat{i}}\right). 
\end{align}
Now we make an analytical continuation of $\overset{(0)}{S}(\mathcal{L}_2)$ to arbitrary $d$. Note that both sides of (\ref{sect 2: get SL2}) are defined with respective to $\overset{(0)}{g}_{ij}$. Thus, we have obtained a non-perturbative entropy formula for arbitrary metrics $\overset{(0)}{g}_{ij}$.  We verify that $\overset{(0)}{S}(\mathcal{L}_2)$ agrees with the HEE of curvature-squared gravity obtained in \cite{Dong:2013qoa, Camps:2013zua}, which is a non-trivial test of our approach.  Now we have completed the demonstration of our RHEE approach. In the next section, we will use it to derive $S(\mathcal{L}_3)$, whose calculations are more complicated.

\section{HEE for cubic HD gravity}

This section derives $S(\mathcal{L}_3)$ by applying the RHEE approach developed in sect.2. To do so, we need to work out $\sqrt{H}$ (\ref{sect 2:H}), $\sqrt{h}$ (\ref{sect 2:h}) and $S\left(\mathcal{L}_2\right)$ (\ref{sect 2: derive SL2}) to the next order. Let us discuss them one by one below. Those not interested in complex calculations can skip to the final result  (\ref{sect 3:covariant entropy for L3}) and various verifications in sect. 4. 

\subsection{Expansion of metric determinants}

Following \cite{Dong:2013qoa}, we expand the metric $\overset{(0)}{g}_{ij}$ near the boundary entangling surface $\gamma$
\begin{align}\label{sect 3:conical metric}
	ds^2&=\overset{(0)}{g}_{ij}dx^idx^j\notag\\
	&=dzd\bar{z}+T(\bar{z}dz-zd\bar{z})^2+2i U_{\hat{i}}(\bar{z}dz-zd\bar{z})dy^{\hat{i}}+(h_{\hat{i}\hat{j}}-2k_{a\hat{i}\hat{j}}x^a+Q_{ab\hat{i}\hat{j}}x^ax^b+...)dy^{\hat{i}}dy^{\hat{j}},
\end{align}
where $(a,b)$ denote the indices orthogonal to the entangling surface $\gamma$ located at $x^a=(z,\bar{z})=0$.  Note that the extrinsic curvature $k_{a\hat{i}\hat{j}}$ is related to the one defined in (\ref{sect 2:extrinsic curvature}) by
\begin{align}\label{sect 3: extrinsic curvature}
k_{a\hat{i}\hat{j}}=n^m{}_ah_{\hat{i}}{}^ih_{\hat{j}}{}^j\overset{(0)}{k}_{mij},
\end{align}
where $x^i=(x^a,y^{\hat{i}} )$, $n^i{}_a=\frac{\partial x^i}{\partial x^a}$ are the normal vectors orthogonal to the entangling surface, and 
$h^{i}{}_{\hat{i}}=\frac{\partial x^i }{\partial y^{\hat{i}}}$ are projections tangent to the entangling surface. Below, we first take the metric (\ref{sect 3:conical metric}) to do calculations and then rewrite the results into covariant forms.

From (\ref{sect 2:bulk metric}, \ref{sect 2:FG gauge}, \ref{sect 2:chosen gauge},\ref{sect 2:embedding functions}) and (\ref{sect 3:conical metric}), we can derive the induced metric (\ref{sect 2: induced metric on Gamma}) on bulk entangling surface $\Gamma$. We have for the ``$\tau\tau$" components
\begin{align}\label{sect 3:htautau}
	\hat{h}_{\tau\tau}&=\partial_\tau X^\mu\partial_\tau X^\nu \hat{G}_{\mu\nu}=\frac{1}{4\tau^2}\bigg[1+\tau\overset{(1)}{h}_{\tau\tau}+\tau^2\overset{(2)}{h}_{\tau\tau}+O(\tau^2)\bigg],
\end{align}
where$\overset{(1)}{h}_{\tau\tau}$ is (\ref{sect 2:h1ij}), and $\overset{(2)}{h}_{\tau\tau}$ is given by
\begin{equation}\label{sect 3:h2tautau}
	\overset{(2)}{h}_{\tau\tau}=\frac{1}{(d-2)^2}\overset{(0)}{k}{}^i\overset{(0)}{k}{}^j\overset{(1)}{g}_{ij}+\frac{8}{(d-2)}\overset{(0)}{k}{}^j\overset{(2)}{X}{}^i\overset{(0)}{g}_{ij}.
\end{equation}
Note that we have dropped terms like ``$\overset{(1)}{X}{}^i\overset{(1)}{X}{}^j\overset{(1)}{X}{}^k\partial_k\overset{(0)}{g}{}_{ij}\sim O(x^a)$'', because it vanishes on the boundary entangling surface $\gamma$ $(x^a=0)$ and thus is irrelevant to the entropy $\overset{(0)}{S}(\mathcal{L}_3)$ defined on $\gamma$. Remarkably, as we will show in section.3.3, $\overset{(2)}{X}{}^i$ only appears in a total derivative term of  $\overset{(0)}{S}(\mathcal{L}_3)$. Thus, we can drop it for simplicity. 

Similarly, we get the ``$\hat{i}\hat{j}$" components of the induced metric (\ref{sect 2: induced metric on Gamma})
\begin{align}\label{sect 3:hij}
	\hat{h}_{\hat{i}\hat{j}}=\frac{1}{\rho}\partial_{\hat{i}} X^i\partial_{\hat{j}} X^j g_{ij}=\frac{1}{\tau}(\overset{(0)}{h}_{\hat{i}\hat{j}}+\tau\overset{(1)}{h}_{\hat{i}\hat{j}}+\tau^2\overset{(2)}{h}_{\hat{i}\hat{j}}+...)
\end{align}
where $\overset{(1)}{h}_{\hat{i}\hat{j}}$ is given by (\ref{sect 2:h1ij}) and $\overset{(2)}{h}_{\hat{i}\hat{j}}$ reads
\begin{align}\label{sect 3:h2ij}
	\overset{(2)}{h}_{\hat{i}\hat{j}}=&\overset{(0)}{g}_{mn}(\frac{1}{4(d-2)^2}\partial_{\hat{i}}\overset{(0)}{k}{}^m\partial_{\hat{j}}\overset{(0)}{k}{}^n+\partial_{\hat{i}}\overset{(2)}{X}{}^m\partial_{\hat{j}}\overset{(0)}{X}{}^n+\partial_{\hat{i}}\overset{(0)}{X}{}^m\partial_{\hat{j}}\overset{(2)}{X}{}^n)\notag\\
	&+\frac{1}{2(d-2)}(\partial_{\hat{i}}\overset{(0)}{k}{}^m\overset{(1)}{g}{}_{m\hat{j}}+\partial_{\hat{j}}\overset{(0)}{k}{}^m\overset{(1)}{g}{}_{m\hat{i}})+\frac{1}{2(d-2)^2}\epsilon_{mn}(\partial_{\hat{i}}\overset{(0)}{k}{}^m\overset{(0)}{k}{}^nU_{\hat{j}}+\partial_{\hat{j}}\overset{(0)}{k}{}^m\overset{(0)}{k}{}^nU_{\hat{i}})\notag\\
	&+\overset{(2)}{g}{}_{\hat{i}\hat{j}}+\frac{1}{2(d-2)}\overset{(0)}{k}{}^m\partial_m\overset{(1)}{g}{}_{\hat{i}\hat{j}}+\frac{1}{4(d-2)^2}\overset{(0)}{k}{}^a\overset{(0)}{k}{}^bQ_{ab\hat{i}\hat{j}}+\partial_{\hat{i}} \overset{(0)}{X}{}^m \partial_{\hat{j}} \overset{(0)}{X}{}^n \overset{(2)}{X}{}^k\partial_k\overset{(0)}{g}{}_{mn}\notag\\
	=&\overset{(0)}{h}{}^{i}{}_{\hat{i}}\overset{(0)}{h}{}^{j}{}_{\hat{j}}\Big[\frac{1}{4(d-2)^2}(\nabla_i\overset{(0)}{k}{}^m\nabla_j\overset{(0)}{k}{}^n\overset{(0)}{g}_{mn}-\overset{(0)}{k}{}^m\overset{(0)}{k}{}^nR_{minj})+\frac{1}{(d-2)}\nabla_{(i}\overset{(0)}{k}{}^m\overset{(1)}{g}_{j)m}\notag\\
	&+\frac{1}{2(d-2)}\overset{(0)}{k}{}^m\nabla_m\overset{(1)}{g}_{ij}+\overset{(2)}{g}_{ij} \Big]+\overset{(0)}{g}_{mn}(\partial_{\hat{i}}\overset{(2)}{X}{}^m\partial_{\hat{j}}\overset{(0)}{X}{}^n+\partial_{\hat{i}}\overset{(0)}{X}{}^m\partial_{\hat{j}}\overset{(2)}{X}{}^n)+\partial_{\hat{i}} \overset{(0)}{X}{}^m \partial_{\hat{j}} \overset{(0)}{X}{}^n \overset{(2)}{X}{}^k\partial_k\overset{(0)}{g}{}_{mn}.
\end{align}
Above, we have rewrite (\ref{sect 3:h2ij}) into covariant expression in the second equation. It agrees with the results of \cite{Miao:2015iba} for $d=6$. Here we focus on the general dimensions. 

From the above equations, we finally obtain the metric determinants to the next order
\begin{equation}\label{sect 3:H}
	\sqrt{H}=\frac{\sqrt{\overset{(0)}{h}}}{2\tau^{\frac{d}{2}}}\bigg[1+\frac{\tau}{2}(\overset{(1)}{h}_{\tau\tau}+\overset{(1)}{h}_{\hat{i}}{}^{\hat{i}})+\frac{\tau^2}{4}(2\overset{(2)}{h}_{\tau\tau}-\frac{1}{2}\overset{(1)}{h}_{\tau\tau}{}^2+2\overset{(2)}{h}_{\hat{i}}{}^{\hat{i}}-\overset{(1)}{h}_{\hat{i}\hat{j}}\overset{(1)}{h}{}^{\hat{i}\hat{j}}+\frac{1}{2}\overset{(1)}{h}_{\hat{i}}{}^{\hat{i}}{}^2+\overset{(1)}{h}_{\tau\tau}\overset{(1)}{h}_{\hat{i}}{}^{\hat{i}})+O(\tau^3)\bigg],
\end{equation}
\begin{equation}\label{sect 3:h}
	\sqrt{h}=\frac{\sqrt{\overset{(0)}{h}}}{\tau^{\frac{d}{2}-1}}\bigg[1+\frac{\tau}{2}\overset{(1)}{h}_{\hat{i}}{}^{\hat{i}}+\frac{\tau^2}{4}(2\overset{(2)}{h}_{\hat{i}}{}^{\hat{i}}-\overset{(1)}{h}_{\hat{i}\hat{j}}\overset{(1)}{h}{}^{\hat{i}\hat{j}}+\frac{1}{2}\overset{(1)}{h}_{\hat{i}}{}^{\hat{i}}{}^2)+O(\tau^3)\bigg],
\end{equation}
where various $\overset{(0)}{h}$ are given by (\ref{sect 2:h1ij},\ref{sect 3:h2tautau},\ref{sect 3:h2ij}).

\subsection{Expansion of  $S\left(\mathcal{L}_2\right)$}

Let us go on to study the expansions of $S\left(\mathcal{L}_2\right)$. Recall that we have derived $S(\mathcal{L}_2(\overset{(0)}{g}))$ (\ref{sect 2: get SL2}) for general metric $\overset{(0)}{g}_{ij}$ in sect.2.3. From (\ref{sect 2: get SL2}), we get
\begin{align} \label{sect 3: SL2}
S(\mathcal{L}_2(\tilde{g}))=2 \pi  \left(\frac{d-2}{d-1}\tilde{R} -k^ik_i-2 \tilde{R}^{\hat{i}}{}_{\hat{i}}\right),
\end{align}
for metric $\tilde{g}_{ij}=g_{ij}/\rho=\overset{(0)}{g}_{ij}/\rho+...$ (\ref{sect 2:bulk metric}). After complicated calculations, we derive 
\begin{align} \label{sect 3: SL2 expansion}
S(\mathcal{L}_2(\tilde{g}))= \tau\ 2 \pi   \left(\frac{d-2}{d-1}\overset{(0)}{R} -\overset{(0)}{k}{}^i\overset{(0)}{k}{}_i-2 \overset{(0)}{R}{}^{\hat{i}}{}_{\hat{i}}\right)+\tau^2 \overset{(1)}{S}(\mathcal{L}_2) +O(\tau^3),
\end{align}
with $\overset{(1)}{S}(\mathcal{L}_2)$ given by
\begin{align} \label{sect 3: SL2 expansion 1}
 \overset{(1)}{S}(\mathcal{L}_2)=&2\pi\Big[2\overset{(1)}{h}{}^{ij}\overset{(0)}{R}_{ij}
	-2\overset{(0)}{h}{}^{ij}\overset{(1)}{R}_{ij}
	+2\overset{(1)}{h}{}^{\hat{i}\hat{j}}\overset{(0)}{k}{}^i{}_{\hat{i}\hat{j}}\overset{(0)}{k}{}_i
	-2\overset{(0)}{h}{}^{\hat{i}\hat{j}}\overset{(1)}{k}{}^i{}_{\hat{i}\hat{j}}\overset{(0)}{k}{}_i
	-\overset{(1)}{g}_{ij}\overset{(0)}{k}{}^i\overset{(0)}{k}{}^j
	-\overset{(1)}{X}{}^m\partial_m\overset{(0)}{g}{}_{ij}\overset{(0)}{k}{}^i\overset{(0)}{k}{}^j\notag\\
	&-\frac{d-2}{(d-1)}\overset{(1)}{g}{}^{ij}\overset{(0)}{R}{}_{ij}+\frac{d-2}{(d-1)}\overset{(1)}{X}{}^m\partial_m\overset{(0)}{g}{}^{ij}\overset{(0)}{R}{}_{ij}+\frac{d-2}{(d-1)}\overset{(0)}{g}{}^{ij}\overset{(1)}{R}{}_{ij}\Big],
\end{align}
where $\overset{(1)}{X}{}^m$ is (\ref{sect 2:X(1)}),  $\overset{(1)}{k}$ and $\overset{(1)}{R}$ are defined with respect to $g_{ij}$ below. We have
\begin{equation} \label{sect 3: formula1}
	\gamma{}^{\hat{i}}_{\hat{j}\hat{k}}=\overset{(0)}{\gamma}{}^{\hat{i}}_{\hat{j}\hat{k}}+	\tau\overset{(1)}{\gamma}{}^{\hat{i}}_{\hat{j}\hat{k}}+\dots
\end{equation}
\begin{equation}\label{sect 3: formula2}
	\Gamma^i_{jk}=\overset{(0)}{\Gamma}{}^i_{jk}+	\tau\overset{(1)}{\Gamma}{}^i_{jk}+\dots
\end{equation}
\begin{equation} \label{sect 3: formula3}
	R_{ij}=\overset{(0)}{R}{}_{ij}+	\tau\overset{(1)}{R}_{ij}+\dots
\end{equation}
\begin{equation} \label{sect 3: formula4}
	k^i{}_{\hat{i}\hat{j}}=\overset{(0)}{k}{}^i{}_{\hat{i}\hat{j}}+	\tau\overset{(1)}{k}{}^i{}_{\hat{i}\hat{j}}+\dots
\end{equation}
with
\begin{equation} \label{sect 3: formula5}
	\overset{(1)}{\gamma}{}^{\hat{i}}_{\hat{j}\hat{k}}=\frac{1}{2}\overset{(0)}{h}{}^{\hat{i}\hat{l}}(D_{\hat{j}}\overset{(1)}{h}{}_{\hat{l}\hat{k}}
	+D_{\hat{k}}\overset{(1)}{h}{}_{\hat{l}\hat{j}}
	-D_{\hat{l}}\overset{(1)}{h}{}_{\hat{j}\hat{k}}),
\end{equation}
\begin{equation} \label{sect 3: formula6}
	\overset{(1)}{\Gamma}{}^i_{jk}=\frac{1}{2}\overset{(0)}{g}{}^{il}\big[\nabla_{j}(\overset{(1)}{g}{}_{lk}+\overset{(1)}{X}{}^m\partial_m\overset{(0)}{g}{}_{lk})
	+\nabla_{k}(\overset{(1)}{g}{}_{lj}+\overset{(1)}{X}{}^m\partial_m\overset{(0)}{g}{}_{lj})
	-\nabla_{l}(\overset{(1)}{g}{}_{jk}+\overset{(1)}{X}{}^m\partial_m\overset{(0)}{g}{}_{jk})\big],
\end{equation}
\begin{align} \label{sect 3: formula7}
	\overset{(1)}{R}{}_{ij}=&\frac{1}{2}\overset{(0)}{g}{}^{kl}\nabla_k\big[\nabla_{i}(\overset{(1)}{g}{}_{lj}+\overset{(1)}{X}{}^m\partial_m\overset{(0)}{g}{}_{lj})
	+\nabla_{j}(\overset{(1)}{g}{}_{li}+\overset{(1)}{X}{}^m\partial_m\overset{(0)}{g}{}_{li})
	-\nabla_{l}(\overset{(1)}{g}{}_{ij}+\overset{(1)}{X}{}^m\partial_m\overset{(0)}{g}{}_{ij})\big]\notag\\
	&-\frac{1}{2}\overset{(0)}{g}{}^{kl}\nabla_j\nabla_{i}(\overset{(1)}{g}{}_{kl}+\overset{(1)}{X}{}^m\partial_m\overset{(0)}{g}{}_{kl}),
\end{align}
and 
\begin{align} \label{sect 3: formula8}
	\overset{(1)}{k}{}^i{}_{\hat{i}\hat{j}}=\partial_{\hat{i}}\partial_{\hat{j}}\overset{(1)}{X}{}^i
	-\overset{(0)}{\gamma}{}^{\hat{l}}_{\hat{i}\hat{j}}\partial_{\hat{l}}\overset{(1)}{X}{}^i
	-\overset{(1)}{\gamma}{}^{\hat{l}}_{\hat{i}\hat{j}}\partial_{\hat{l}}\overset{(0)}{X}{}^i
	+2\overset{(0)}{\Gamma}{}^i_{jk}\partial_{\hat{i}}\overset{(1)}{X}{}^j\partial_{\hat{j}}\overset{(0)}{X}{}^k
	+\overset{(1)}{\Gamma}{}^i_{jk}\partial_{\hat{i}}\overset{(0)}{X}{}^j\partial_{\hat{j}}\overset{(0)}{X}{}^k,
\end{align}
where $D_{\hat{i}}$ is the covariant derivative with respect to $\overset{(0)}{h}{}_{\hat{i}\hat{j}}$. Recall that $\tilde{g}_{ij}=g_{ij}/\rho$. Thus, we have $\tilde{R}_{ij}=R_{ij}$ and
\begin{align}\label{sect 3:R expansion}
	\tilde{R}=&\tilde{g}^{ij}\tilde{R}_{ij}=\tau g^{ij}R_{ij}\notag\\
	=&\tau\big[\overset{(0)}{g}{}^{ij}-\tau(\overset{(1)}{g}{}^{ij}-\overset{(1)}{X}{}^m\partial_m\overset{(0)}{g}{}^{ij})\big](\overset{(0)}{R}{}_{ij}+	\tau\overset{(1)}{R}_{ij})\notag\\
	=&\tau\overset{(0)}{R}
	+\tau^2\big[-\overset{(1)}{g}{}^{ij}\overset{(0)}{R}{}_{ij}+\overset{(1)}{X}{}^m\partial_m\overset{(0)}{g}{}^{ij}\overset{(0)}{R}{}_{ij}+\overset{(0)}{g}{}^{ij}\overset{(1)}{R}{}_{ij}\big]\notag\\
	=&\tau\overset{(0)}{R}+\tau^2\big[-\overset{(1)}{g}{}^{ij}\overset{(0)}{R}{}_{ij}+\overset{(1)}{X}{}^m\partial_m\overset{(0)}{g}{}^{ij}\overset{(0)}{R}{}_{ij}+\nabla^i\nabla^j\overset{(1)}{g}{}_{ij}-\Box\overset{(1)}{g}{}^{i}{}_i\notag\\
	&+\nabla^i\nabla^j(\overset{(1)}{X}{}^m\partial_m\overset{(0)}{g}{}_{ij})-\Box(\overset{(0)}{g}{}^{kl}\overset{(1)}{X}{}^m\partial_m\overset{(0)}{g}{}_{kl})\big],
\end{align}
\begin{align}\label{sect 3:hijRij expansion}
	\hat{h}^{\hat{i}\hat{j}}\tilde{R}_{\hat{i}\hat{j}}=&\tau(\overset{(0)}{h}{}^{\hat{i}\hat{j}}-\tau\overset{(1)}{h}{}^{\hat{i}\hat{j}})(\overset{(0)}{R}{}_{\hat{i}\hat{j}}+	\tau\overset{(1)}{R}_{\hat{i}\hat{j}})\notag\\
	=&\tau\overset{(0)}{h}{}^{\hat{i}\hat{j}}\overset{(0)}{R}{}_{\hat{i}\hat{j}}+\tau^2\big[-\overset{(1)}{h}{}^{\hat{i}\hat{j}}\overset{(0)}{R}{}_{\hat{i}\hat{j}}+\overset{(0)}{h}{}^{ij}\overset{(1)}{R}{}_{ij}\big]\notag\\
	=&\tau\overset{(0)}{h}{}^{\hat{i}\hat{j}}\overset{(0)}{R}{}_{\hat{i}\hat{j}}+\tau^2\big[-\overset{(1)}{h}{}^{\hat{i}\hat{j}}\overset{(0)}{R}{}_{\hat{i}\hat{j}}+\overset{(0)}{h}{}^{ij}\nabla^k\nabla_i\overset{(1)}{g}{}_{kj}-\frac{1}{2}\overset{(0)}{h}{}^{ij}\Box\overset{(1)}{g}{}_{ij}\notag\\
	&-\frac{1}{2}\overset{(0)}{h}{}^{ij}\nabla_j\nabla_i\overset{(1)}{g}{}^k{}_{k}+\overset{(0)}{h}{}^{ij}\nabla^k\nabla_i(\overset{(1)}{X}{}^m\partial_m\overset{(0)}{g}{}_{kj})-\frac{1}{2}\overset{(0)}{h}{}^{ij}\Box(\overset{(1)}{X}{}^m\partial_m\overset{(0)}{g}{}_{ij})\notag\\
	&-\frac{1}{2}\overset{(0)}{h}{}^{ij}\nabla_j\nabla_i(\overset{(0)}{g}{}^{kl}\overset{(1)}{X}{}^m\partial_m\overset{(0)}{g}{}_{kl})\big],
\end{align}
\begin{align}\label{sect 3:k expansion}
	k^i=\hat{h}^{\hat{i}\hat{j}}k^i{}_{\hat{i}\hat{j}}=\tau(\overset{(0)}{h}{}^{\hat{i}\hat{j}}-\tau\overset{(1)}{h}{}^{\hat{i}\hat{j}})(\overset{(0)}{k}{}^i{}_{\hat{i}\hat{j}}+	\tau\overset{(1)}{k}{}^i{}_{\hat{i}\hat{j}})
	=\tau\overset{(0)}{k}{}^i+\tau^2(-\overset{(1)}{h}{}^{\hat{i}\hat{j}}\overset{(0)}{k}{}^i{}_{\hat{i}\hat{j}}+\overset{(0)}{h}{}^{\hat{i}\hat{j}}\overset{(1)}{k}{}^i{}_{\hat{i}\hat{j}}),
\end{align}
\begin{align}\label{sect 3:kk expansion}
	k^ik_i=&\tilde{g}_{ij}k^ik^j=\frac{1}{\tau}\big[\overset{(0)}{g}{}_{ij}+\tau(\overset{(1)}{g}{}_{ij}+\overset{(1)}{X}{}^m\partial_m\overset{(0)}{g}{}_{ij})\big]\big[\tau\overset{(0)}{k}{}^i+\tau^2(-\overset{(1)}{h}{}^{\hat{i}\hat{j}}\overset{(0)}{k}{}^i{}_{\hat{i}\hat{j}}+\overset{(0)}{h}{}^{\hat{i}\hat{j}}\overset{(1)}{k}{}^i{}_{\hat{i}\hat{j}})\big]\notag\\
	&\big[\tau\overset{(0)}{k}{}^j+\tau^2(-\overset{(1)}{h}{}^{\hat{m}\hat{n}}\overset{(0)}{k}{}^j{}_{\hat{m}\hat{n}}+\overset{(0)}{h}{}^{\hat{m}\hat{n}}\overset{(1)}{k}{}^j{}_{\hat{m}\hat{n}})\big]\notag\\
	=&\tau\overset{(0)}{k}{}^i\overset{(0)}{k}{}_i
	+\tau^2(-2\overset{(1)}{h}{}^{\hat{i}\hat{j}}\overset{(0)}{k}{}^i{}_{\hat{i}\hat{j}}\overset{(0)}{k}{}_i+2\overset{(0)}{h}{}^{\hat{i}\hat{j}}\overset{(1)}{k}{}^i{}_{\hat{i}\hat{j}}\overset{(0)}{k}{}_i
+\overset{(1)}{g}_{ij}\overset{(0)}{k}{}^i\overset{(0)}{k}{}^j+\overset{(1)}{X}{}^m\partial_m\overset{(0)}{g}{}_{ij}\overset{(0)}{k}{}^i\overset{(0)}{k}{}^j).
\end{align}
Substituting (\ref{sect 3:R expansion},\ref{sect 3:hijRij expansion},\ref{sect 3:kk expansion}) into (\ref{sect 3: SL2}), we finally obtain  $\overset{(1)}{S}(\mathcal{L}_2)$ (\ref{sect 3: SL2 expansion 1}).

\subsection{Expression of $S\left(\mathcal{L}_3\right)$}

Now, we are ready to derive $S\left(\mathcal{L}_3\right)$. Since $\tilde{g}_{ij}=\overset{(0)}{g}_{ij}/\rho+...$, we have  $S\left(\mathcal{L}_3\right)\sim \tilde{R}^2\sim \rho^2 \overset{(0)}{R}{}^2$, which yields
\begin{align}\label{sect 3: S3 expansion}
S\left(\mathcal{L}_3\right)=\tau^2 \overset{(0)}{S}\left(\mathcal{L}_3\right)+O(\tau^3),
\end{align}
where $ \overset{(0)}{S}\left(\mathcal{L}_3\right)= S(\mathcal{L}_3(\overset{(0)}{g}_{ij}))$. Substituting (\ref{sect 3:H},\ref{sect 3:h},\ref{sect 3: SL2 expansion}) and (\ref{sect 3: S3 expansion}) into (\ref{sect 2: RHEE}), we can derive the UV divergent term of RHEE up to $1/\epsilon^{d-6}$. As we have shown in sect. 2, the $1/\epsilon^{d-4}$ terms of RHEE cancel. Thus, we focus on the $1/\epsilon^{d-6}$ terms below. We have
\begin{align}\label{sect 3:second order term of entropy}
S_{\text{RHEE}}=&\pi\int_{\epsilon^2} d\tau \int_\gamma d^{d-2}y\sqrt{\overset{(0)}{h}}\frac{1}{2\tau^{\frac{d}{2}-2}}(2\overset{(2)}{h}_{\tau\tau}-\frac{1}{2}\overset{(1)}{h}_{\tau\tau}{}^2+2\overset{(2)}{h}_{\hat{i}}{}^{\hat{i}}-\overset{(1)}{h}_{\hat{i}\hat{j}}\overset{(1)}{h}{}^{\hat{i}\hat{j}}+\frac{1}{2}\overset{(1)}{h}_{\hat{i}}{}^{\hat{i}}{}^2+\overset{(1)}{h}_{\tau\tau}\overset{(1)}{h}_{\hat{i}}{}^{\hat{i}})\notag\\
	&+\int_{\gamma (\tau=\epsilon^2)} d^{d-2}y\sqrt{\overset{(0)}{h}}\frac{1}{\tau^{\frac{d}{2}-1}}\bigg[1+\frac{\tau}{2}\overset{(1)}{h}_{\hat{i}}{}^{\hat{i}}+\frac{\tau^2}{4}(2\overset{(2)}{h}_{\hat{i}}{}^{\hat{i}}-\overset{(1)}{h}_{\hat{i}\hat{j}}\overset{(1)}{h}{}^{\hat{i}\hat{j}}+\frac{1}{2}\overset{(1)}{h}_{\hat{i}}{}^{\hat{i}}{}^2)\bigg]\notag\\
	&\bigg[-\frac{4\pi}{d-2}-\frac{4\pi\tau}{(d-4)(d-2)^2}(-\overset{(0)}{h}{}^{ij}\overset{(0)}{R}_{ij}-\frac{1}{2}\overset{(0)}{k}{}^m\overset{(0)}{k}{}_m+\frac{d-2}{2(d-1)}\overset{(0)}{R})\notag\\
	&-\frac{4\pi\tau^2}{(d-4)(d-2)^2}(\overset{(1)}{h}{}^{ij}\overset{(0)}{R}_{ij}-\overset{(0)}{h}{}^{ij}\overset{(1)}{R}_{ij}+\overset{(1)}{h}{}^{\hat{i}\hat{j}}\overset{(0)}{k}{}^i{}_{\hat{i}\hat{j}}\overset{(0)}{k}{}_i-\overset{(0)}{h}{}^{\hat{i}\hat{j}}\overset{(1)}{k}{}^i{}_{\hat{i}\hat{j}}\overset{(0)}{k}{}_i-\frac{1}{2}\overset{(1)}{g}_{ij}\overset{(0)}{k}{}^i\overset{(0)}{k}{}^j-\frac{1}{2}\overset{(1)}{X}{}^m\partial_m\overset{(0)}{g}{}_{ij}\overset{(0)}{k}{}^i\overset{(0)}{k}{}^j\notag\\
	&-\frac{d-2}{2(d-1)}\overset{(1)}{g}{}^{ij}\overset{(0)}{R}{}_{ij}+\frac{d-2}{2(d-1)}\overset{(1)}{X}{}^m\partial_m\overset{(0)}{g}{}^{ij}\overset{(0)}{R}{}_{ij}+\frac{d-2}{2(d-1)}\overset{(0)}{g}{}^{ij}\overset{(1)}{R}{}_{ij})
	+\frac{2\tau^2}{(d-6)(d-4)(d-2)^3}\overset{(0)}{S}(\mathcal{L}_3)\bigg]\notag\\
	=&\int_{\gamma(\tau=\epsilon^2)} d^{d-2}y\sqrt{\overset{(0)}{h}}\frac{1}{\epsilon^{d-6}}\Big[\frac{\pi}{(d-6)}(2\overset{(2)}{h}_{\tau\tau}-\frac{1}{2}\overset{(1)}{h}_{\tau\tau}{}^2+2\overset{(2)}{h}_{\hat{i}}{}^{\hat{i}}-\overset{(1)}{h}_{\hat{i}\hat{j}}\overset{(1)}{h}{}^{\hat{i}\hat{j}}+\frac{1}{2}\overset{(1)}{h}_{\hat{i}}{}^{\hat{i}}{}^2+\overset{(1)}{h}_{\tau\tau}\overset{(1)}{h}_{\hat{i}}{}^{\hat{i}})\notag\\
	&-\frac{\pi}{(d-2)}(2\overset{(2)}{h}_{\hat{i}}{}^{\hat{i}}-\overset{(1)}{h}_{\hat{i}\hat{j}}\overset{(1)}{h}{}^{\hat{i}\hat{j}}+\frac{1}{2}\overset{(1)}{h}_{\hat{i}}{}^{\hat{i}}{}^2)
	-\frac{2\pi}{(d-4)(d-2)^2}\overset{(1)}{h}_{\hat{i}}{}^{\hat{i}}(-\overset{(0)}{h}{}^{ij}\overset{(0)}{R}_{ij}-\frac{1}{2}\overset{(0)}{k}{}^m\overset{(0)}{k}{}_m+\frac{d-2}{2(d-1)}\overset{(0)}{R})\notag\\
	&-\frac{4\pi}{(d-4)(d-2)^2}(\overset{(1)}{h}{}^{ij}\overset{(0)}{R}_{ij}-\overset{(0)}{h}{}^{ij}\overset{(1)}{R}_{ij}+\overset{(1)}{h}{}^{\hat{i}\hat{j}}\overset{(0)}{k}{}^i{}_{\hat{i}\hat{j}}\overset{(0)}{k}{}_i-\overset{(0)}{h}{}^{\hat{i}\hat{j}}\overset{(1)}{k}{}^i{}_{\hat{i}\hat{j}}\overset{(0)}{k}{}_i-\frac{1}{2}\overset{(1)}{g}_{ij}\overset{(0)}{k}{}^i\overset{(0)}{k}{}^j-\frac{1}{2}\overset{(1)}{X}{}^m\partial_m\overset{(0)}{g}{}_{ij}\overset{(0)}{k}{}^i\overset{(0)}{k}{}^j\notag\\
	&-\frac{d-2}{2(d-1)}\overset{(1)}{g}{}^{ij}\overset{(0)}{R}{}_{ij}+\frac{d-2}{2(d-1)}\overset{(1)}{X}{}^m\partial_m\overset{(0)}{g}{}^{ij}\overset{(0)}{R}{}_{ij}+\frac{d-2}{2(d-1)}\overset{(0)}{g}{}^{ij}\overset{(1)}{R}{}_{ij})+\frac{2}{(d-6)(d-4)(d-2)^3}\overset{(0)}{S}(\mathcal{L}_3)
	\Big].
\end{align}
For sufficiently large dimensions ($d>6$), the finiteness of RHEE (\ref{sect 3:second order term of entropy}) yields
\begin{align}\label{sect 3:entropy for L3}
	S(\mathcal{L}_3)=&\pi\left\{-4(d-2)^2(d-4)\overset{(2)}{h}_{\hat{i}}{}^{\hat{i}}
	+2(d-2)^2(d-4)\overset{(1)}{g}_{\hat{i}\hat{j}}\overset{(1)}{g}{}^{\hat{i}\hat{j}}
	-2(d-2)^2(\overset{(1)}{g}_{\hat{i}}{}^{\hat{i}})^2\right.\notag\\
	&-4(d-2)(d-4)\overset{(1)}{g}{}^{\hat{i}\hat{j}}k^nk_{n\hat{i}\hat{j}}
	+3(d-2)k^ik_i\overset{(1)}{g}_{\hat{i}}{}^{\hat{i}}
	-(d-2)(d-4)k^ik^j\overset{(1)}{g}_{ij}\notag\\
	&+2(d-4)k^nk_{n\hat{i}\hat{j}}k_mk^{m\hat{i}\hat{j}}
	-\frac{3d-4}{4(d-2)}(k^ik_i)^2
	-8(d-2)^2(d-4)\overset{(2)}{X}{}^ik_i\notag\\
	&+2(d-2)(d-6)\bigg[\overset{(1)}{g}{}^{\hat{i}\hat{j}}R_{\hat{i}\hat{j}}
	-\frac{1}{(d-2)}k_mk^{m\hat{i}\hat{j}}R_{\hat{i}\hat{j}}
	-h^{ij}\nabla^k\nabla_i\overset{(1)}{g}{}_{kj}
	+\frac{1}{2}h^{ij}\Box\overset{(1)}{g}{}_{ij}\notag\\
	&+\frac{1}{2}h^{ij}\nabla_j\nabla_i\overset{(1)}{g}{}^k{}_{k}
	-\frac{1}{2(d-2)}h^{ij}\nabla^k\nabla_i(k^m\partial_mg_{kj})
	+\frac{1}{4(d-2)}h^{ij}\Box(k^m\partial_mg_{ij})\notag\\
	&+\frac{1}{4(d-2)}h^{ij}\nabla_j\nabla_i(g^{kl}k^m\partial_mg_{kl})
	+\overset{(1)}{g}{}^{\hat{i}\hat{j}}k^i{}_{\hat{i}\hat{j}}k_i
	-\frac{1}{d-2}k_ik^i{}_{\hat{i}\hat{j}}k_jk^{j\hat{i}\hat{j}}
	-\frac{1}{2}\overset{(1)}{g}_{ij}k^ik^j
	\notag\\
	&-\frac{1}{2(d-2)}h^{\hat{i}\hat{j}}k_i\partial_{\hat{i}}\partial_{\hat{j}}k^i
	+\frac{1}{2(d-2)}h^{\hat{i}\hat{j}}k_i\gamma^{\hat{l}}_{\hat{i}\hat{j}}\partial_{\hat{l}}k^i
	-\frac{1}{(d-2)}h^{lk}k_i\Gamma^i_{jk}\partial_{l}k^j\notag\\
	&-\frac{1}{2}h^{jk}k^l(\nabla_{j}\overset{(1)}{g}{}_{lk}
	+\nabla_{k}\overset{(1)}{g}{}_{lj}
	-\nabla_{l}\overset{(1)}{g}{}_{jk})
	-\frac{1}{4(d-2)}h^{jk}k^l\big[\nabla_{j}(k^m\partial_mg_{lk})\notag\\
	&+\nabla_{k}(k^m\partial_mg_{lj})
	-\nabla_{l}(k^m\partial_mg_{jk})\big]
	-\frac{d-2}{2(d-1)}\overset{(1)}{g}{}^{ij}R_{ij}
	+\frac{1}{4(d-1)}k^m\partial_mg^{ij}R_{ij}\notag\\
	&+\frac{d-2}{2(d-1)}\nabla^i\nabla^j\overset{(1)}{g}{}_{ij}-\frac{d-2}{2(d-1)}\Box\overset{(1)}{g}{}^{i}{}_i+\frac{1}{4(d-1)}\nabla^i\nabla^j(k^m\partial_mg_{ij})\notag\\
	&\left.-\frac{1}{4(d-1)}\Box(g^{kl}k^m\partial_mg_{kl})
	\bigg]\right\},
\end{align}
where $\nabla_i$ are the covariant derivatives with respect to $\overset{(0)}{g}{}_{ij}$, $ \overset{(0)}{h}{}_{\hat{i}\hat{j}}=\frac{\partial \overset{(0)}{X}{}^i}{\partial y^{\hat{i}}} \frac{\partial \overset{(0)}{X}{}^j}{\partial y^{\hat{j}}} \overset{(0)}{g}{}_{ij}$, $ \overset{(0)}{h}{}^{\hat{i}\hat{j}} $ is the inverse metric of  $\overset{(0)}{h}{}_{\hat{i}\hat{j}}$, $ \overset{(0)}{h}{}^{ij}=\frac{\partial \overset{(0)}{X}{}^i}{\partial y^{\hat{i}}} \frac{\partial \overset{(0)}{X}{}^j}{\partial y^{\hat{j}}}  \overset{(0)}{h}{}^{\hat{i}\hat{j}} $, the expressions of $\overset{(1)}{g},\overset{(2)}{g}$ and $\overset{(2)}{h}$ can be found in (\ref{sect 2:g1ij}, \ref{sect 2:g2ij}, \ref{sect 3:h2ij}). For simplicity, we have ignored ${}^{(0)}$ in the above equation. Besides, 
 we have used $\overset{(1)}{X}{}^m\partial_m\overset{(0)}{g}{}_{ij}k^ik^j=0$ on the entangling surface $\gamma$.  Note that $k^m$ of $\nabla (k^m \partial_m g)$ and $\nabla^2 (k^m \partial_m g)$ comes from $ \overset{(1)}{X}{}^{i}(y^{\hat{i}})\sim\overset{(0)}{k}{}^i(y^{\hat{j}})$ (\ref{sect 2:X(1)}) of eqs.(\ref{sect 3: formula6},\ref{sect 3: formula7}), which is independent of the normal coordinates $x^a=(z,\bar{z})$.  As a result, the normal derivatives of extrinsic curvatures $\partial_a k^m= \partial_a \partial_b k^m=0$ vanish in $\nabla (k^m \partial_m g)$ and $\nabla^2 (k^m \partial_m g)$ of (\ref{sect 3:entropy for L3}). It is different from the convention that $\partial_a k_{b \hat{i}\hat{j}}=-Q_{ab \hat{i}\hat{j}} \ne 0$. To avoid confusion, we carefully rewrite the entropy (\ref{sect 3:entropy for L3}) into an covariant form (\ref{sect 3:covariant entropy for L3}) without normal derivatives of $k^m $. 
 
 Let us try to simplify the complicated non-covariant entropy formula (\ref{sect 3:entropy for L3}).  First, we show that  the $\overset{(2)}{X}{}^{i}$ term in (\ref{sect 3:entropy for L3}) is a total derivative. Note that $\overset{(2)}{h}{}^{\hat{i}}{}_{\hat{i}}$ (\ref{sect 3:h2ij}) also contains $\overset{(2)}{X}{}^{i}$. Selecting all the $\overset{(2)}{X}{}^{i}$ term in (\ref{sect 3:entropy for L3}) and integrating on the entangling surface, we get a total derivative term
\begin{align}\label{sect 3:X(2) term of entropy for L3}
	S_{\overset{(2)}{X}}=&-8(d-2)^2(d-4)\pi\int_\gamma d^{d-2}y\sqrt{\overset{(0)}{h}}\big[
	(\overset{(0)}{h}{}^{\hat{i}\hat{j}}\partial_{\hat{i}}\overset{(0)}{X}{}^{m}\partial_{\hat{j}}\overset{(2)}{X}{}^{n}\overset{(0)}{g}_{mn}+\frac{1}{2}\overset{(0)}{h}{}^{mn}\overset{(2)}{X}{}^{k}\partial_k\overset{(0)}{g}_{mn})
	+\overset{(2)}{X}{}^{i}\overset{(0)}{k}{}^j\overset{(0)}{g}_{ij}\big]\notag\\
	=&-8(d-2)^2(d-4)\pi\int_\gamma d^{d-2}y\sqrt{\overset{(0)}{h}}\bigg[\overset{(2)}{X}{}^{i}\overset{(0)}{g}_{ij}\overset{(0)}{h}{}^{\hat{m}\hat{n}}(\overset{(0)}{k}{}^j_{\hat{m}\hat{n}}-\partial_{\hat{m}}\partial_{\hat{n}}\overset{(0)}{X}{}^{j}+\overset{(0)}{\gamma}{}^{\hat{l}}{}_{\hat{m}\hat{n}}\partial_{\hat{l}}\overset{(0)}{X}{}^{j}-\overset{(0)}{\Gamma}{}^{j}{}_{kl}\partial_{\hat{m}}\overset{(0)}{X}{}^{k}\partial_{\hat{n}}\overset{(0)}{X}{}^{l})\bigg]\notag\\
	&-8(d-2)^2(d-4)\pi\int_\gamma d^{d-2}y\partial_{\hat{j}}(\sqrt{\overset{(0)}{h}}\overset{(0)}{h}{}^{\hat{i}\hat{j}}\partial_{\hat{i}}\overset{(0)}{X}{}^{m}\overset{(2)}{X}{}^{n}\overset{(0)}{g}_{mn})\notag\\
	=&-8(d-2)^2(d-4)\pi\int_\gamma d^{d-2}y\sqrt{\overset{(0)}{h}}D_{\hat{i}}\overset{(2)}{X}{}^{\hat{i}}.
\end{align}
Therefore we can drop $\overset{(2)}{X}$ in (\ref{sect 3:entropy for L3}) safely on the closed entangling surfaces, such as the spherical black hole horizon. As for the open entangling surfaces, our entropy formula holds up to some total derivatives. 

Removing $\overset{(2)}{X}$ terms and separating the non-covariant term, we further simplify the entropy (\ref{sect 3:entropy for L3}) as
\begin{align}\label{sect 3:simplified entropy for L3}
	S(\mathcal{L}_3)=&\pi\left\{-4(d-2)^2(d-4)\overset{(2)}{h}_{\hat{i}}{}^{\hat{i}}
	+2(d-2)^2(d-4)\overset{(1)}{g}_{\hat{i}\hat{j}}\overset{(1)}{g}{}^{\hat{i}\hat{j}}
	-2(d-2)^2(\overset{(1)}{g}_{\hat{i}}{}^{\hat{i}})^2\right.\notag\\
	&-4(d-2)(d-4)\overset{(1)}{g}{}^{\hat{i}\hat{j}}k^nk_{n\hat{i}\hat{j}}
	+3(d-2)k^ik_i\overset{(1)}{g}_{\hat{i}}{}^{\hat{i}}
	-(d-2)(d-4)k^ik^j\overset{(1)}{g}_{ij}\notag\\
	&+2(d-4)k^nk_{n\hat{i}\hat{j}}k_mk^{m\hat{i}\hat{j}}
	-\frac{3d-4}{4(d-2)}(k^ik_i)^2
	\notag\\
	&+2(d-2)(d-6)\bigg[\overset{(1)}{g}{}^{\hat{i}\hat{j}}R_{\hat{i}\hat{j}}
	-\frac{1}{(d-2)}k_mk^{m\hat{i}\hat{j}}R_{\hat{i}\hat{j}}
	-h^{ij}\nabla^k\nabla_i\overset{(1)}{g}{}_{kj}
	+\frac{1}{2}h^{ij}\Box\overset{(1)}{g}{}_{ij}\notag\\
	&+\frac{1}{2}h^{ij}\nabla_j\nabla_i\overset{(1)}{g}{}^k{}_{k}
	+\overset{(1)}{g}{}^{\hat{i}\hat{j}}k^i{}_{\hat{i}\hat{j}}k_i
	-\frac{1}{d-2}k_ik^i{}_{\hat{i}\hat{j}}k_jk^{j\hat{i}\hat{j}}
	-\frac{1}{2}\overset{(1)}{g}_{ij}k^ik^j
	\notag\\
	&\left.-\frac{1}{2}h^{jk}k^l(2\nabla_{k}\overset{(1)}{g}{}_{lj}
	-\nabla_{l}\overset{(1)}{g}{}_{jk})
	-\frac{d-2}{2(d-1)}\overset{(1)}{g}{}^{ij}R_{ij}
	+\frac{d-2}{2(d-1)}\nabla^i\nabla^j\overset{(1)}{g}{}_{ij}
	-\frac{d-2}{2(d-1)}\Box\overset{(1)}{g}{}^{i}{}_i\bigg] \right\}\notag\\
&+S(\mathcal{L}_3)_{\text{non-cov}},
\end{align}
with
\begin{align}\label{sect 3:non-covariant term of entropy for L3}
	S(\mathcal{L}_3)_{\text{non-cov}}=\pi(d-6)\bigg[&-h^{ij}\nabla^k\nabla_i(k^m\partial_mg_{kj})
	+\frac{1}{2}h^{ij}\Box(k^m\partial_mg_{ij})
	+\frac{1}{2}h^{ij}\nabla_j\nabla_i(g^{kl}k^m\partial_mg_{kl})\notag\\
	&
	-h^{\hat{i}\hat{j}}k_i\partial_{\hat{i}}\partial_{\hat{j}}k^i
	+h^{\hat{i}\hat{j}}k_i\gamma^{\hat{l}}_{\hat{i}\hat{j}}\partial_{\hat{l}}k^i
	-2h^{lk}k_i\Gamma^i_{jk}\partial_{l}k^j
	-h^{jk}k^l\nabla_{j}(k^m\partial_mg_{lk})\notag\\
	&
	+\frac{1}{2}h^{jk}k^l\nabla_{l}(k^m\partial_mg_{jk})
	-\frac{(d-2)}{2(d-1)}k^m\partial_mg_{ij}R^{ij}
	+\frac{(d-2)}{2(d-1)}\nabla^i\nabla^j(k^m\partial_mg_{ij})\notag\\
	&
	-\frac{(d-2)}{2(d-1)}\Box(g^{kl}k^m\partial_mg_{kl})
	\bigg].
\end{align} 
Recall again that the above $k^m$ of $\nabla (k^m \partial_m g)$ and $\nabla^2 (k^m \partial_m g)$ are independent of the normal coordinates $x^a=(z,\bar{z})$. We first calculate the non-covariant term (\ref{sect 3:non-covariant term of entropy for L3}) with the conical metric (\ref{sect 3:conical metric}), and then take values at the entangling surface $\gamma$ with $x^a=0$. Finally, we rewrite the results into covariant form. In this way, we get 
\begin{align}\label{sect 3:simplify non-covariant term of entropy for L3 }
	S(\mathcal{L}_3)_{\text{non-cov}}=\pi(d-6)\bigg[&-\frac{d-2}{d-1}k^mk^{n\hat{i}\hat{j}}R_{m\hat{i}n\hat{j}}
	-\frac{1}{d-1}k^mk^nR_{m\hat{i}n}{}^{\hat{i}}
	-\frac{2}{d-1}k_mk^{m\hat{i}\hat{j}}R^b{}_{\hat{i}b\hat{j}}
	\notag\\
	&
	+\frac{2}{d-1}k_mk^{m\hat{i}\hat{j}}R_{\hat{i}\hat{j}}
	+\frac{3d-2}{3(d-1)}D^{\hat{i}}k^mR^a{}_{ma\hat{i}}
	-\frac{d}{d-1}h^{ij}\nabla_ik^mR^a{}_{maj}
	\notag\\
	&
	+\frac{d-2}{d-1}D^{\hat{i}}k^mR_{m\hat{i}}
	-\frac{d-2}{d-1}h^{ij}\nabla_{i}k^mR_{mj}
	-\frac{1}{d-1}h^{ij}k^m\nabla_mR_{ij}
	\notag\\
	&
	+\frac{d-2}{2(d-1)}k^mn^{ij}n^{kl}\nabla_mR_{ikjl}
	-h^{ij}k^m\nabla_i(h^l{}_j\nabla_lk_m)
	+h^{i}{}_{\hat{i}}\nabla_{i}k_mD^{\hat{i}}k^m
	\notag\\
	&
	-D_{\hat{i}}k_mD^{\hat{i}}k^m
	+\frac{2}{d-1}h^{ij}h^{kl}k^m\nabla_k(h^n{}_l\nabla_nk_{mij})
	-\frac{2}{d-1}h^{ij}h^{kl}k^m\nabla_k(h^n{}_i\nabla_nk_{mlj})
	\notag\\
	&
	-\frac{d}{d-1}D^{\hat{i}}(k^mR_{m\hat{j}\hat{i}}{}^{\hat{j}})
	+\frac{d}{d-1}D^{\hat{i}}(k_{m\hat{i}\hat{j}}D^{\hat{j}}k^m)
	-\frac{d}{d-1}D^{\hat{i}}(k_{m}D_{\hat{i}}k^m)
	\bigg].
\end{align}
where $n_{ij}=g_{ij}-h_{ij}$. 
The last three terms are total derivative terms, which we drop for simplicity. 

Substituting (\ref{sect 3:simplify non-covariant term of entropy for L3 }) into (\ref{sect 3:simplified entropy for L3}), we finally obtain the covariant expression of $S(\mathcal{L}_3)$
\begin{align}\label{sect 3:covariant entropy for L3}
	S(\mathcal{L}_3)=&\pi\left\{-4(d-2)^2(d-4)\overset{(2)}{h}_{\hat{i}}{}^{\hat{i}}
	+2(d-2)^2(d-4)\overset{(1)}{g}_{\hat{i}\hat{j}}\overset{(1)}{g}{}^{\hat{i}\hat{j}}
	-2(d-2)^2(\overset{(1)}{g}_{\hat{i}}{}^{\hat{i}})^2\right.\notag\\
	&-2(d-2)^2\overset{(1)}{g}{}^{\hat{i}\hat{j}}k^nk_{n\hat{i}\hat{j}}
	+3(d-2)k^ik_i\overset{(1)}{g}_{\hat{i}}{}^{\hat{i}}
	-2(d-2)(d-5)k^ik^j\overset{(1)}{g}_{ij}\notag\\
	&+4k^nk_{n\hat{i}\hat{j}}k_mk^{m\hat{i}\hat{j}}
	-\frac{3d-4}{4(d-2)}(k^ik_i)^2
	\notag\\
	&+2(d-2)(d-6)\bigg[\overset{(1)}{g}{}^{\hat{i}\hat{j}}R_{\hat{i}\hat{j}}
	-h^{ij}\nabla^k\nabla_i\overset{(1)}{g}{}_{kj}
	+\frac{1}{2}h^{ij}\Box\overset{(1)}{g}{}_{ij}
	+\frac{1}{2}h^{ij}\nabla_j\nabla_i\overset{(1)}{g}{}^k{}_{k}
	\notag\\
	&-\frac{1}{2}h^{jk}k^l(2\nabla_{k}\overset{(1)}{g}{}_{lj}
	-\nabla_{l}\overset{(1)}{g}{}_{jk})
	-\frac{d-2}{2(d-1)}\overset{(1)}{g}{}^{ij}R_{ij}
	+\frac{d-2}{2(d-1)}\nabla^i\nabla^j\overset{(1)}{g}{}_{ij}
	-\frac{d-2}{2(d-1)}\Box\overset{(1)}{g}{}^{i}{}_i\bigg]\notag\\
	&+(d-6)\bigg[-\frac{d-2}{d-1}k^mk^{n\hat{i}\hat{j}}R_{m\hat{i}n\hat{j}}
	-\frac{1}{d-1}k^mk^nR_{m\hat{i}n}{}^{\hat{i}}
	-\frac{2}{d-1}k_mk^{m\hat{i}\hat{j}}R^b{}_{\hat{i}b\hat{j}}
	-\frac{2(d-2)}{d-1}k_mk^{m\hat{i}\hat{j}}R_{\hat{i}\hat{j}}
	\notag\\
	&
	+\frac{3d-2}{3(d-1)}D^{\hat{i}}k^mR^a{}_{ma\hat{i}}
	-\frac{d}{d-1}h^{ij}\nabla_ik^mR^a{}_{maj}
	+\frac{d-2}{d-1}D^{\hat{i}}k^mR_{m\hat{i}}
	-\frac{d-2}{d-1}h^{ij}\nabla_{i}k^mR_{mj}
	\notag\\
	&
	-\frac{1}{d-1}h^{ij}k^m\nabla_mR_{ij}
	+\frac{d-2}{2(d-1)}k^mn^{ij}n^{kl}\nabla_mR_{ikjl}
	+h^{i}{}_{\hat{i}}\nabla_{i}k_mD^{\hat{i}}k^m
	-D_{\hat{i}}k_mD^{\hat{i}}k^m
	\notag\\
	&
	\left.
	-h^{ij}k^m\nabla_i(h^l{}_j\nabla_lk_m)
	+\frac{2}{d-1}h^{ij}h^{kl}k^m\nabla_k(h^n{}_l\nabla_nk_{mij})
	-\frac{2}{d-1}h^{ij}h^{kl}k^m\nabla_k(h^n{}_i\nabla_nk_{mlj})
	\bigg]\right\},
\end{align}
where $\overset{(1)}{g}$ and $\overset{(2)}{h}$ are given by (\ref{sect 2:g1ij},\ref{sect 3:h2ij}),  $n_{ij}=g_{ij}-h_{ij}$ with $h_{ij}$ the induced metric on the entangling surface $\gamma$, $\overset{(0)}{X}{}^i=(x^a, y^{\hat{i}})$ with $a$ and $\hat{i}$ labeling the normal and tangent directions of the entangling surface, respectively. Recall again that we have ignored $\overset{(0)}{}$ above. Eq.(\ref{sect 3:covariant entropy for L3}) is the main result of this paper. It should be mentioned that one author of this paper had derived $S(\mathcal{L}_3)$ for $d=6$ from the universal term of HEE \cite{Miao:2015iba}. Here, we reproduce the special case of \cite{Miao:2015iba} by applying RHEE method, and obtain non-trivial corrections for $d\ne 6$. We will test the entropy formula (\ref{sect 3:covariant entropy for L3}) in the next section.

\section{Tests of cubic entropy formula}

This section tests the generalized gravitational entropy $S(\mathcal{L}_3)$ (\ref{sect 3:covariant entropy for L3}). First, we verify that it agrees with the Wald entropy \cite{Wald:1993nt} on the Killing horizon with $k=U=Q=0$. Then we confirm that it is consistent with the HEE for general HD gravity \cite{Miao:2014nxa} with $k=U=0, Q\ne 0$. Next, we check our calculations by considering flat metrics with non-zero extrinsic curvatures $k\ne 0, Q\sim k^2 \ne 0, U=0$. Finally, we show that our entropy formula $S(\mathcal{L}_3)$ is consistent with HEE under field redefinition. Below, let us discuss them in order.

\subsection{Wald entropy}

We first consider Wald entropy \cite{Wald:1993nt} on the entangling surface with $k_{a\hat{i}\hat{j}}=U_{\hat{i}}=Q_{ab\hat{i}\hat{j}}=0$. See the conical metric (\ref{sect 3:conical metric})
 for the definitions of $k, U,Q$.  For zero extrinsic curvatures $k=0$, our entropy (\ref{sect 3:covariant entropy for L3}) becomes
\begin{align}\label{sect 4: S(L3) with zero extrinsic curvature}
	S(\mathcal{L}_3)=&\pi\left\{-4(d-2)^2(d-4)\overset{(2)}{g}_{\hat{i}}{}^{\hat{i}}
	+2(d-2)^2(d-4)\overset{(1)}{g}_{\hat{i}\hat{j}}\overset{(1)}{g}{}^{\hat{i}\hat{j}}
	-2(d-2)^2(\overset{(1)}{g}_{\hat{i}}{}^{\hat{i}})^2\right.\notag\\
	&+2(d-2)(d-6)\bigg[\overset{(1)}{g}{}^{\hat{i}\hat{j}}R_{\hat{i}\hat{j}}
	-\frac{d-2}{2(d-1)}\overset{(1)}{g}{}^{ij}R_{ij}
	-h^{ij}\nabla^k\nabla_i\overset{(1)}{g}{}_{kj}
	+\frac{1}{2}h^{ij}\Box\overset{(1)}{g}{}_{ij}\notag\\
	&\left.+\frac{1}{2}h^{ij}\nabla_j\nabla_i\overset{(1)}{g}{}^k{}_{k}
	+\frac{d-2}{2(d-1)}\nabla^i\nabla^j\overset{(1)}{g}{}_{ij}
	-\frac{d-2}{2(d-1)}\Box\overset{(1)}{g}{}^{i}{}_i\bigg]\right\}\notag\\
	=& \pi\big[\frac{3d+2}{d-1}R^{ij}R_{ij}
	-\frac{3d(d+2)}{4(d-1)^2}R^2
	+\frac{3d+2}{d-1}RR^{a}{}_{a}
	-4R^{ai}R_{ai}
	-8R^{ij}R^{a}{}_{iaj}
	+4R^{ab}R_{ab}\notag\\
	&-2(R^{a}{}_{a})^2
	+\frac{4}{d-1}\Box R+\frac{2(d-2)}{d-1}n^{ij}\nabla_i\nabla_{j}R-4n^{ij}\Box R_{ij}\big].
\end{align}

On the Killing horizon (entangling surface with $k=U=Q=0$), the entropy of general HD gravity is given by the famous Wald formula \cite{Wald:1993nt} 
\begin{align}\label{sect 4: SWald}
	S_{\text{Wald}}=2\pi \int d^{d-2}y \sqrt{h}\frac{\delta L}{\delta R_{ijkl}} \epsilon_{ij}\epsilon_{kl}.
\end{align}
By applying this formula, we derive the Wald entropy for $\mathcal{L}_3$ (\ref{sect 2: L3}) with the metric $\overset{(0)}{g}_{ij}$
\begin{align}\label{sect 4: SWald(L3)}
	S(\mathcal{L}_3)_{\text{Wald}}=&\pi\big[\frac{3d+2}{d-1}R^{ij}R_{ij}+\frac{3d+2}{d-1}RR^a{}_a-\frac{3d(d+2)}{4(d-1)^2}R^2-4(R^a{}_a)^2\notag\\
	&+4R_{ab}R^{ab}-8R^{ij}R^a{}_{iaj}+\frac{2d}{d-1}\Box R-4n^{ij}\Box R_{ij}\big]
\end{align}
where we have ignored $\overset{(0)}{}$ for simplicity. From (\ref{sect 4: S(L3) with zero extrinsic curvature}) and (\ref{sect 4: SWald(L3)}), we derive the difference in entropy 
\begin{align}\label{sect 4: entropy difference}
	\Delta S&= S(\mathcal{L}_3)-S(\mathcal{L}_3)_{\text{Wald}}\notag\\
	&=  \pi\big[
	-4R^{ai}R_{ai}+2(R^{a}{}_{a})^2-\frac{2(d-2)}{d-1}h^{ij}\nabla_i\nabla_{j}R\big]
\end{align}
For the conical metric (\ref{sect 3:conical metric}) with $k=U=0$, we derive
\begin{align}\label{sect 4: entropy difference1}
	\Delta S= -32\pi Q_{zz}Q_{\bar{z}\bar{z}},
\end{align}
which vanishes on the Killing horizon with $k=U=Q=0$. Thus, our results agree with the Wald entropy. However, there is an entropy mismatch even for zero extrinsic curvatures $k=U=0, Q\ne 0$. This is the so-called Hung-Myers-Smolkin puzzle \cite{Hung:2011xb}. It can be resolved by considering the anomalous contributions to the entropy \cite{Miao:2014nxa}. We will discuss it in the next subsection.

\subsection{HEE with zero extrinsic curvatures}

For HD gravity with one derivative of curvature, the HEE formula is given by \cite{Miao:2014nxa} for $k=U=0$, \footnote{In the notation of this paper, the entropy density of $R$ is $4\pi$, which differs by a minus sign from that of \cite{Miao:2014nxa}.}
\begin{align}\label{sect 4: general HEE formula}
	S_{\text{HEE}}=-2\pi \int d^{d-2}y \sqrt{h}\Big[\frac{\delta L}{\delta R_{z\bar{z}z\bar{z}}} +64 \Big(\frac{\partial^2 L}{\partial \nabla_z R_{zizj} \partial \nabla_{\bar{z}} R_{\bar{z}k\bar{z}l}} \Big)_{\alpha_1} \frac{Q_{zzij}Q_{\bar{z}\bar{z}kl}}{\beta_{\alpha_1}}\Big],
\end{align}
where the first term is Wald entropy and the second term is an anomalous term from the derivative of curvatures. For the metric (\ref{sect 3:conical metric}), the anomalous entropy of $\mathcal{L}_3$ (\ref{sect 2: L3}) reads
\begin{align}\label{sect 4: anomalous entropy}
	S_{\text{ano}}=-32\pi Q_{zz}Q_{\bar{z}\bar{z}},
\end{align}
which is exactly the entropy mismatch (\ref{sect 4: entropy difference1}). Now we resolve the entropy mismatch, and verify that our entropy (\ref{sect 3:covariant entropy for L3}) derived from RHEE method agrees with the one \cite{Miao:2014nxa} by using squashed-cone method for $k=U=0$. We remark that the splitting problem disappears for $k=0$ \cite{Miao:2014nxa,Miao:2015iba}. We discuss the case with $k\ne 0$ in the next subsection.

\subsection{Flat space with non-zero extrinsic curvatures}

Let us consider the more interesting case with non-zero extrinsic curvatures. For simplicity, we focus on the flat space with $\overset{(1)}{g}_{ij}=\overset{(2)}{g}_{ij}=\overset{(0)}{R}_{ijkl}=0$. Note that $\overset{(0)}{R}_{ijkl}=0$ imposes constraints on $k,T,U,Q$. We further focus on constant extrinsic curvatures, i.e.,  $D_{\hat{i}}k^n{}_{\hat{j}\hat{k}}=0$. Then, our entropy (\ref{sect 3:covariant entropy for L3}) reduces to
\begin{align}\label{sect 4:S(L3) in flat spacetime}
	S(\mathcal{L}_3)=&\pi\left\{-(d-4)g_{mn}h^{ij}\nabla_{i}k^m\nabla_{j}k^n
	+4k^mk_{m\hat{i}\hat{j}}k_nk^{n\hat{i}\hat{j}}
	-\frac{3d-4}{4(d-2)}(k^mk_m)^2\right.\notag\\
	&+(d-6)\big[
	-h^{ij}k^m\nabla_i(h^l{}_j\nabla_lk_m)
	+\frac{2}{d-1}h^{ij}h^{kl}k^m\nabla_k(h^n{}_l\nabla_nk_{mij})\notag\\
	&\left.
	-\frac{2}{d-1}h^{ij}h^{kl}k^m\nabla_k(h^n{}_i\nabla_nk_{mlj})
	\big]\right\}.
\end{align}
We remark that the splitting problem appears in the $k^4$ terms. Thus, it is necessary to make more tests of the $k^4$ terms for our entropy formula (\ref{sect 3:covariant entropy for L3}). 

The following AdS metric in bulk satisfies all the above requirements,
\begin{align}\label{sect 4: bulk metric test}
ds^2=\frac{dz^2-dt^2+dr^2+r^2d\Omega_m^2+\sum_{i=1}^{d-2-m} dy_i^2 }{z^2},
\end{align}
where $d\Omega_m^2$ denotes the metric of $m$-dimensional unite sphere.  Note that we have $\rho=z^2$ in the FG gauge (\ref{sect 2:bulk metric}). Then, it is clear that  $\overset{(1)}{g}_{ij}=\overset{(2)}{g}_{ij}=0$ for the above metric.  We choose the boundary entangling surface $\gamma$ at $t=0, r=r_0$. Then the entropy (\ref{sect 4:S(L3) in flat spacetime}) becomes 
\begin{align}\label{sect 4:S(L3)0}
\overset{(0)}{S}(\mathcal{L}_3)&=\frac{\pi  (4-3 d)}{4 (d-2)} k^4+ 2 \pi k^2 \text{Tr}(k^2) \notag\\
	&=-\frac{\pi  m^3 (3 d m-8 d-4 m+16)}{4 (d-2) r_0^4},
\end{align}
where we have used $k=-m/r_0,  \text{Tr}(k^2)=m/r_0^2$ to derive the second line of the above equation. Below, we re-derive (\ref{sect 4:S(L3)0}) as a double check of our calculations.

We assume the embedding function of bulk entangling surface $\Gamma$ as
\begin{align}\label{sect 4: bulk entangling surface1}
& t=0,\\
& r=r(z)=r_0+ a_1 z^2+ a_2 z^4+a_3 z^6+... \label{sect 4: bulk entangling surface2}
\end{align}
where $a_n$ are constants corresponding to $\overset{(n)}{X}$ of (\ref{sect 2:embedding functions}). The bulk entropy functional is given by Ryu-Takayanagi formula \cite{Ryu:2006bv}
\begin{align} \label{sect 4: bulk entropy}
	S_{\text{bulk}}=4\pi \int_\Gamma d^{d-1}Y\sqrt{H}=4\pi \int_{\epsilon} dz \frac{r(z)^m\sqrt{1+r'(z)^2}}{z^{d-1}},
\end{align}
where we have set the volumes of $m$-dimensional unite sphere and tangential space $\int d^{d-2-m}y$ to be one. Taking the variation of (\ref{sect 4: bulk entropy}), we derive the Euler-Lagrange equation 
\begin{align} \label{sect 4: Euler-Lagrange equation}
r(z) \left((d-1) r'(z)^3+(d-1) r'(z)-z r''(z)\right)+m z \left(r'(z)^2+1\right)=0.
\end{align}
Solving it perturbatively, we get
\begin{align} \label{sect 4: a1}
a_1=\frac{m}{2(2-d) r_0},
\end{align}
which agrees with (\ref{sect 2:X(1)}) with $k=-m/r_0$. Here we derive $a_1$ from the Euler-Lagrange equation instead of the PBH transformation. Thus it is a test of our calculations based PBH transformation in sect.2 and sect.3.  Similarly, we can solve $a_2$ at the next order of $z^2$. However, we leave $a_2$ free  to test if $a_2=\overset{(2)}{X}$ is irrelevant to $S(\mathcal{L}_3)$, as we stated in sect.3.  Substituting (\ref{sect 4: bulk entangling surface2}) with (\ref{sect 4: a1}) into (\ref{sect 4: bulk entropy}), we derive the bulk entropy 
\begin{align} \label{sect 4: bulk entropy expansion}
	S_{\text{bulk}}&=\frac{4\pi r_0^m }{\epsilon^{d-2}} \Big[ \frac{1}{d-2}
-\frac{(d-3) m^2}{2 (d-4) (d-2)^2 } \frac{\epsilon^2}{r_0^2} \notag\\
&+\frac{m   \left(8 a_2 (d-6) (d-2)^3 r_0^3+\left(d^2-6 d+7\right) m^3-(d-2)^2 m^2\right)}{8 (d-6) (d-2)^4 }  \frac{\epsilon^4}{r_0^4} +O(\epsilon^6)\Big].
\end{align}

Let us go on the study the boundary entropy in (\ref{sect 2: RHEE}).  The induced metric on AdS boundary $M$ and boundary entangling surface $\gamma$ read
\begin{align}\label{sect 4: metric on M}
ds_M^2&=\tilde{g}_{ij} dx^i dx^j=\frac{-dt^2+dr^2+r^2d\Omega_m^2+\sum_{i=1}^{d-2-m} dy_i^2 }{\epsilon^2}, \\
ds_{\gamma}^2&=\hat{h}_{\hat{i}\hat{j}} dy^{\hat{i}} dy^{\hat{j}}=\frac{r(\epsilon)^2d\Omega_m^2+\sum_{i=1}^{d-2-m} dy_i^2 }{\epsilon^2}. \label{sect 4: metric on gamma}
\end{align}
From (\ref{sect 4: bulk entangling surface2}, \ref{sect 4: a1}, \ref{sect 4: metric on gamma}), we derive
\begin{align} \label{sect 4: h expansion}
\sqrt{h}=\frac{r(\epsilon)^m}{\epsilon^{d-2}}=\frac{r_0^m}{\epsilon^{d-2}}\Big[ 1+\frac{m^2  }{4-2 d}\frac{\epsilon^2}{r_0^2}+\frac{1}{8} m  \left(8 a_2 r_0^3+\frac{(m-1) m^2}{(d-2)^2}\right) \frac{\epsilon^4}{r_0^4}+O\left(\epsilon^6\right) \Big].
\end{align}
From (\ref{sect 4: metric on M}) with $t=0, r=r(\epsilon)$, we derive $\tilde{R}=\tilde{R}^{\hat{i}}_{\ \hat{i}}=0$ and $k^i k_i=m^2\epsilon^2/r(\epsilon)^2$, which together with (\ref{sect 3: SL2}) yields
\begin{align} \label{sect 4: SL2}
S(\mathcal{L}_2)=-2 \pi  k^ik_i=-\frac{2 \pi  m^2 \epsilon ^2}{r_0^2}-\frac{2 \pi  m^3 \epsilon ^4}{(d-2) r_0^4}+O\left(\epsilon ^6\right).
\end{align}
Substituting (\ref{sect 4: h expansion},\ref{sect 4: SL2}) and $S(\mathcal{L}_3)=\epsilon^4 \overset{(0)}{S}(\mathcal{L}_3)+O(\epsilon^6)$ into the second line of (\ref{sect 2: RHEE}), we derive the boundary entropy
\begin{align} \label{sect 4: boundary entropy}
	S_{\text{bdy}}&=\frac{4\pi r_0^m }{\epsilon^{d-2}} \Big[ \frac{-1}{d-2}+\frac{(d-3) m^2 \epsilon ^2}{2 (d-4) (d-2)^2 r_0^2} \notag\\
	&+\frac{\epsilon ^4 \left(-\frac{8 a_2 (d-2)^2 m}{r_0}+\frac{m^3 (d (-m)+d+2 m)}{(d-4) r_0^4}+\frac{4 \overset{(0)}{S}(\mathcal{L}_3)}{\pi  (d-6) (d-4)}\right)}{8 (d-2)^3}+O\left(\epsilon ^5\right) \Big].
\end{align}
From (\ref{sect 4: bulk entropy expansion}) and (\ref{sect 4: boundary entropy}), we finally obtain the total renormalized entropy 
\begin{align} \label{sect 4: total entropy}
	S_{\text{RHEE}}&=S_{\text{bulk}}+S_{\text{bdy}} \notag\\
	&=\frac{r_0^{m-4} \left(\pi  m^3 (3 d m-8 d-4 m+16)+4 (d-2) r_0^4 \overset{(0)}{S}(\mathcal{L}_3)\right)}{\left(2 (d-6) (d-4) (d-2)^4\right) \epsilon ^{d-6}}+O(\frac{1}{\epsilon ^{d-8}}).
\end{align}
As expected, $a_2$ of $S_{\text{bulk}}$ and $S_{\text{bdy}}$ cancel, which support our above statement that $a_2=\overset{(2)}{X}$ is irrelevant to $S(\mathcal{L}_3)$.  The finiteness of (\ref{sect 4: total entropy}) can derive $ \overset{(0)}{S}(\mathcal{L}_3)$, which exactly agrees with (\ref{sect 4:S(L3)0}). Now we finish the tests of ${S}(\mathcal{L}_3)$ for flat space with non-zero extrinsic curvatures.  The calculations of this subsection are much simpler than those of sect.3. Thus it is easier for readers to follow our work.

\subsection{HEE under field redefinition}

In the framework of effective field theory (EFT), one class of HD gravity is perturbatively equivalent to Einstein gravity under suitable field redefinition. For instance, we can rewrite the Lagrangian of Einstein gravity as $\mathcal{L}=R+\lambda_1 R^2+\lambda_2R^{ij}R_{ij}$ by the field redefinition $g_{ij} \to g_{ij}+c_1 R+c_2 R_{ij}$ with small couplings $\lambda_1, \lambda_2, c_1, c_2$. By utilizing the dynamic equations and extreme conditions, one can show the entropy functionals of the two Lagrangians are equivalent. See \cite{Dong:2023bax} for an example. This section shows that the cubic BWHD gravity $\mathcal{L}_3$ is equivalent to Einstein gravity in the sense of EFT. Furthermore, its on-shell entropy is consistent that of Einstein gravity. We remark that,  unlike the RHEE method, the EFT approach cannot derive the complete expression of the entropy functional with extrinsic curvatures for HD gravity, since it has imposed the extreme condition $k^m=0$. Besides, the EFT approach works well only for small HD couplings, while there is no such limitation for the RHEE method.  

We start with the Einstein-Hilbert action with a cosmological constant
\begin{equation} \label{sect 4: Einstein action}
	\tilde{I}=\int d^{d}x \sqrt{|\tilde{g}|}(\tilde{R}-2\Lambda),
\end{equation}
which yields the equations of motion (EOM)
\begin{align}\label{sect 4:Eoms of Einstein action}
	\tilde{E}^{ij}=\frac{1}{\sqrt{\tilde{g}}}\frac{\delta\tilde{I}}{\delta\tilde{g}_{ij}}=-\tilde{R}^{ij}+\frac{1}{2}\tilde{R}\tilde{g}^{ij}-\Lambda\tilde{g}^{ij}.
\end{align}
The HEE of Einstein gravity is given by the Ryu-Takayanagi formula \cite{Ryu:2006bv}
\begin{equation}\label{sect 4: S for Einstein action}
	\tilde{S}=4\pi\int d^{d-2}y\sqrt{\tilde{h}}.
\end{equation}

To generate the $\mathcal{L}_3$ term, we perform the following field redefinition 
\begin{equation}\label{sect 4: field redefinition}
	\tilde{g}_{ij}=g_{ij}-\frac{3(d-2)}{8(d-1)^2}\lambda R^2g_{ij}+\frac{d-6}{4(d-1)}\lambda RR_{ij}+2\lambda R^{kl}R_{ikjl }-\frac{1}{2(d-1)}\lambda\Box Rg_{ij}+\lambda\Box R_{ij}.
\end{equation}
At the linear order of $\lambda$, the action (\ref{sect 4: Einstein action}) becomes 
\begin{align} \label{sect 4: new action under field redefinition}
	I=&\int d^dx \sqrt{g}(R-2\Lambda)+\int d^dx \sqrt{g}E^{ij}\delta g_{ij}\notag\\
	=&\int d^dx \sqrt{g}\big[R-2\Lambda+(-R^{ij}+\frac{1}{2}Rg^{ij}-\Lambda g^{ij})(-\frac{3(d-2)}{8(d-1)^2}\lambda R^2g_{ij}+\frac{d-6}{4(d-1)}\lambda RR_{ij}\notag\\
	&+2\lambda R^{kl}R_{ikjl }-\frac{1}{2(d-1)}\lambda\Box Rg_{ij}+\lambda\Box R_{ij})\notag\\
	=&\int d^dx \sqrt{g}\big[R-2\Lambda+\frac{d^2+8d-12}{8(d-1)^2}\lambda\Lambda R^2
	-2\lambda\Lambda R^{ij}R_{ij}
	+\lambda(\frac{3d+2}{4(d-1)}RR^{ij}R_{ij}\notag\\
	&-\frac{d(d+2)}{16(d-1)^2}R^3-2R^{kl}R_{ikjl}R^{ij}+\frac{d}{4(d-1)}R\Box R-R^{ij}\Box R_{ij})
	\big]\notag\\
	=&\int d^dx \sqrt{g}\big[R-2\Lambda+\frac{d^2+8d-12}{8(d-1)^2}\lambda\Lambda R^2-2\lambda\Lambda R^{ij}R_{ij}+\lambda\mathcal{L}_3
	\big],
\end{align}
where we have dropped some total derivative terms in the above calculations. By applying the HEE formula for curvature-squared gravity \cite{Dong:2013qoa,Camps:2013zua} together with our result $S(\mathcal{L}_3)$ (\ref{sect 3:covariant entropy for L3}), we obtain the HEE for new action (\ref{sect 4: new action under field redefinition})
\begin{align}\label{sect 4: S for new action}
	S&=4\pi\int d^{d-2}y\sqrt{h}\big[1+\frac{d^2+8d-12}{4(d-1)^2}\lambda\Lambda R
	-2\lambda\Lambda (R^{a}_{\ a}-\frac{1}{2}k^ak_a)+\frac{\lambda}{4\pi}S(\mathcal{L}_3)\big].
\end{align}

Now let us compare the two entropies (\ref{sect 4: S for Einstein action}) and (\ref{sect 4: S for new action}). First, expanding the entropy (\ref{sect 4: S for Einstein action}) to the linear order of field redefinition (\ref{sect 4: field redefinition}), we obtain
\begin{align}\label{sect 4:expand S for Einstein action}
	\tilde{S}&=4\pi\int d^{d-2}y\sqrt{\tilde{h}}=4\pi\int d^{d-2}y\sqrt{h}(1+\frac{1}{2}\delta g^{\hat{i}}_{\hat{i}})\notag\\
	&=4\pi\int d^{d-2}y\sqrt{h}\big[1+\frac{\lambda}{2}(-\frac{3(d-2)^2}{8(d-1)^2} R^2+\frac{d-6}{4(d-1)} RR^{\hat{i}}{}_{\hat{i}}+2R^{kl}R^{\hat{i}}{}_{k\hat{i}l}-\frac{d-2}{2(d-1)}\Box R+g^{ij}_{\parallel}\Box R_{ij})\big].
\end{align}
Imposing Einstein equations, we have $\Box R=\Box R_{ij}=0$ and 
\begin{equation} \label{sect 4: on-shell condition1} 
	R=\frac{2d}{d-2}\Lambda, \quad R_{ij}=\frac{2\Lambda}{d-2} g_{ij},\quad R^{a}{}_a=\frac{4}{d-2}\Lambda, \quad R^{\hat{i}}{}_{\hat{i}}=2\Lambda,
\end{equation}
and 
\begin{align}  \label{sect 4: on-shell condition2}
\overset{(1)}{g}_{ij}=-\frac{\Lambda}{(d-1)(d-2)}g_{ij},\ \overset{(2)}{g}_{ij}=\frac{\Lambda^2}{4(d-1)^2(d-2)^2}g_{ij}.
\end{align}
Substituting the above on-shell equations and the extreme condition $k_a=0$ for Einstein gravity, we simplify the two entropies (\ref{sect 4:expand S for Einstein action}) and (\ref{sect 4: S for new action}) as 
\begin{align} \label{sect 4: finial S1}
	\tilde{S}&=4\pi\int d^{d-2}y\sqrt{h}\big[1+\frac{\lambda}{2}(-\frac{3(d-2)^2}{8(d-1)^2} R^2+\frac{d-6}{4(d-1)} RR^{\hat{i}}{}_{\hat{i}}+2R^{kl}R^{\hat{i}}{}_{k \hat{i}l})\big]\notag\\
	&=4\pi\int d^{d-2}y\sqrt{h}\big[1+\frac{\lambda}{2}\Lambda^2(-\frac{3d^2}{2(d-1)^2} +\frac{d(d-6)}{(d-1)(d-2)} +\frac{8}{d-2})\big]\notag\\
	&=4\pi\int d^{d-2}y\sqrt{h}\big[1-\frac{(d-4)(d-2)}{4(d-1)^2}\lambda\Lambda^2\big],
\end{align}
and
\begin{align} \label{sect 4: finial S2}
	S=&4\pi\int d^{d-2}y\sqrt{h}\big[1+\lambda(\frac{d^2+8d-12}{4(d-1)^2}\Lambda R-2\Lambda R^{a}{}_{a}\notag\\
	&-(d-2)^2(d-4)\overset{(2)}{g}_{\hat{i}}{}^{\hat{i}}
	+\frac{1}{2}(d-2)^2(d-4)\overset{(1)}{g}_{\hat{i}\hat{j}}\overset{(1)}{g}{}^{\hat{i}\hat{j}}
	-\frac{1}{2}(d-2)^2(\overset{(1)}{g}_{\hat{i}}{}^{\hat{i}})^2\notag\\
	&+\frac{1}{2}(d-2)(d-6)\overset{(1)}{g}{}^{\hat{i}\hat{j}}R_{\hat{i}\hat{j}}
	-\frac{(d-2)^2(d-6)}{4(d-1)}\overset{(1)}{g}{}^{ij}R_{ij})
	\big]\notag\\
	=&4\pi\int d^{d-2}y\sqrt{h}\big[1+\lambda\Lambda^2(\frac{d(d^2+8d-12)}{2(d-1)^2(d-2)}-\frac{8}{d-2}\notag\\
	&-\frac{(d-2)(d-4)}{4(d-1)^2}+\frac{(d-2)(d-4)}{2(d-1)^2}-\frac{(d-2)^2}{2(d-1)^2}-\frac{(d-6)}{(d-1)}+\frac{d(d-6)}{2(d-1)^2})\big]\notag\\
	=&4\pi\int d^{d-2}y\sqrt{h}\big[1-\frac{(d-4)(d-2)}{4(d-1)^2}\lambda\Lambda^2\big].
\end{align}
Now, we verify that the two entropies indeed agree with each other. In other words, our entropy formula of $S(\mathcal{L}_3)$ (\ref{sect 3:covariant entropy for L3}) is consistent with that of Einstein gravity in the EFT framework. This is another support for our results. We stress again that the EFT method cannot derive the complete entropy formula of $\mathcal{L}_3$ with non-zero extrinsic curvatures, because it imposes the extreme condition $k_a=0$. For Einstein gravity in vacuum, the extreme condition $k_a=0$ is necessary. However, when there are matter fields, we should consider the quantum extremal surface \cite{Faulkner:2013ana,Engelhardt:2014gca} instead of the minimal surface, which yields non-zero $k_a$ generally. Thus, it is necessary to derive a complete entropy formula of $\mathcal{L}_3$ with non-zero extrinsic curvatures, which is what we have done in this paper.

To summarize, we have checked our entropy formula (\ref{sect 3:covariant entropy for L3}) for cubic BWHD gravity by studying Wald entropy, HEE without/with extrinsic curvatures, and HEE in the EFT framework. 

\section{A glance of HEE for quartic HD gravity}

By applying the RHEE approach, we can derive the HEE for quartic BWHD gravity. The quartic BWHD gravity $\mathcal{L}_4$ is defined as the quartic counter term in the holographic renormalization
\begin{align} \label{sect 5: counterterms}
	I_c&=\int_Md^{d}x\sqrt{|\tilde{g}|} \bigg[-2(d-1)-\frac{\tilde{R}}{d-2}-\frac{\mathcal{L}_2}{(d-4)(d-2)^2}
	+\frac{2\mathcal{L}_3}{(d-6)(d-4)(d-2)^3}\notag\\
	&\ \ \ \ \ \ \ \ \ \ \ \ \ \ \ \ \ \ +\frac{2\mathcal{L}_4}{(d-8)(d-6)(d-4)(d-2)^4}+\dots\bigg],
\end{align}
where $\mathcal{L}_4$ is given by \cite{Bueno:2022log} \footnote{Note that our $\mathcal{L}_4=- (d-8)(d-6)(d-4)(d-2)^4 \mathcal{L}_{(4)}$ of  \cite{Bueno:2022log}.}.  Since it is too complicated, we do not show the exact expression of $\mathcal{L}_4$ here. The HEE of $\mathcal{L}_4$ can be derived from the finiteness of RHEE
\begin{align} \label{sect 5: RHEE}
	S_{\text{RHEE}}=&4\pi \int_\Gamma d^{d-1}Y\sqrt{H}+\int_{\gamma=\partial \Gamma} d^{d-2}y\sqrt{h}\bigg[-\frac{4\pi}{d-2}-\frac{S(\mathcal{L}_2)}{(d-4)(d-2)^2} \notag\\
&+\frac{2S(\mathcal{L}_3)}{(d-6)(d-4)(d-2)^3}+\frac{2S(\mathcal{L}_4)}{(d-8)(d-6)(d-4)(d-2)^4}+...\bigg].
\end{align}
For simplicity, we focus on the HEE $S(\mathcal{L}_4)$ in flat space with $R_{ijkl}=D_i k^{j}_{\hat{k}\hat{l}}=0$.  Then, we can follow the simple approach of sect. 4.3 to derive $S(\mathcal{L}_4)$ in this special case. 

Solving perturbatively the Euler-Lagrange equation (\ref{sect 4: Euler-Lagrange equation}) for bulk entangling surface (\ref{sect 4: bulk entangling surface2}), we derive
\begin{align} \label{sect 5: a2}
a_2=-\frac{m^2 \left(d^2-4 d-2 m+4\right)}{8 (d-4) (d-2)^3 r_0^3}. 
\end{align}
Similar to sect. 4.3, we leave $a_3=\overset{(3)}{X}$ free since it does not contribute to $S(\mathcal{L}_4)$ as we will show below. Substituting (\ref{sect 4: bulk entangling surface2}, \ref{sect 5: a2}) into (\ref{sect 4: bulk entropy}), we derive the bulk contribution to RHEE  at order $\epsilon ^{8-d}$
\begin{align} \label{sect 5: bulk entropy}
	S_{\text{bulk}}|_{\epsilon ^{8-d}}&=\frac{-\pi  r_0^{m-6}}{12 (d-8) (d-4)^2 (d-2)^6} \Big[ -48 a_3 (d-8) (d-4)^2 (d-2)^5 m r_0^5\notag\\
	& +(d (d (d ((d-17) d+115)-407)+748)-536) m^6\notag\\
	&+m^4 \left((d-2)^3 (5 (d-8) d+68)-6 (d-2)^2 (d ((d-12) d+44)-52) m\right) \Big].
\end{align}
Let us go on to discuss the boundary contributions to RHEE.  We find the following formulas are useful
\begin{align} \label{sect 5: bdy terms 1}
&\sqrt{h}=\frac{r(\epsilon)^m}{\epsilon^{d-2}}, \ \ \ \ \ \ \ \ \ S(\mathcal{L}_2)=-2 \pi  k^ik_i=-2\pi m^2\epsilon^2/r(\epsilon)^2, \\
&S(\mathcal{L}_3)=-\frac{\pi  m^3 (3 d m-8 d-4 m+16)\epsilon^4}{4 (d-2) r(\epsilon)^4}, \ S(\mathcal{L}_4)=\epsilon^6 \overset{(0)}{S}(\mathcal{L}_4).  \label{sect 5: bdy terms 2}
\end{align}
The first two formulas are obtained in sect.4.3, $S(\mathcal{L}_3)$ can be derived from $\overset{(0)}{S}(\mathcal{L}_3)$ (\ref{sect 4:S(L3)0}) by considering suitable Weyl rescale and replacing $r_0$ with $r(\epsilon)$. That is because $S(\mathcal{L}_3)$ and $\overset{(0)}{S}(\mathcal{L}_3)$ are defined with respect to the metrics $ds_{\gamma}^2=\frac{r(\epsilon)^2d\Omega_m^2+\sum_{i=1}^{d-2-m} dy_i^2 }{\epsilon^2} $ and $ds_{\gamma_0}^2=r(0)^2d\Omega_m^2+\sum_{i=1}^{d-2-m} dy_i^2 $, respectively.  From (\ref{sect 4: bulk entangling surface2}, \ref{sect 5: a2}) and (\ref{sect 5: bdy terms 1},\ref{sect 5: bdy terms 2}), we derive the boundary contributions to RHEE (\ref{sect 5: RHEE}) at order $\epsilon ^{8-d}$
\begin{align} \label{sect 5: bdy entropy}
	S_{\text{bdy}}|_{\epsilon ^{8-d}}&=\frac{r_0^{m-6}}{12 (d-8) (d-6) (d-4)^2 (d-2)^5} \Big[ 24 \left(d^2-6 d+8\right) r_0^6 \overset{(0)}{S}(\mathcal{L}_4)\notag\\
	&-48 \pi  a_3 (d-4)^2 (d-2)^4 \left(d^2-14 d+48\right) m r_0^5 \notag\\
	& \pi  \left(5 d^5-90 d^4+608 d^3-2136 d^2+4208 d-3456\right) m^4 \notag\\
	&-6 \pi  \left(d^5-20 d^4+152 d^3-568 d^2+1048 d-704\right) m^5\notag\\
	& +\pi  \left(d^5-21 d^4+175 d^3-762 d^2+1756 d-1632\right) m^6 \Big]. 
\end{align}
From (\ref{sect 5: bulk entropy},\ref{sect 5: bdy entropy}) together with the finiteness of $S_{\text{RHEE}}=S_{\text{bulk}}+S_{\text{bdy}}$ for $d>8$, we derive 
\begin{align} \label{sect 5: S4}
\overset{(0)}{S}(\mathcal{L}_4)=\frac{\pi  m^4}{8 (d-4) (d-2)^2 r_0^6}& \Big[ d^2 \left(-30 m^2+288 m-592\right)-16 \left(m^2-20 m+76\right)\notag\\
&+d^3 \left(5 m^2-40 m+72\right)+8 d \left(5 m^2-72 m+188\right) \Big],
\end{align}
which is independent of $a_3$ as we mentioned above. We make analytical continuation of (\ref{sect 5: S4}) to general dimensions expect $d=2, 4$. We can repeat the above approach to derive $S(\mathcal{L}_n)$ with $n\ge 5$. Interestingly, we do not need to know the exact expressions of $\mathcal{L}_n$ in the above approach.

\section{Conclusions and Discussions}

In this section, we develop an approach based on double holography \cite{Chen:2020uac} and renormalized holographic entanglement entropy (RHEE) \cite{Taylor:2016aoi} to derive the holographic entanglement entropy (HEE) for brane-world higher derivative (BWHD) gravity. The BWHD gravities are counterterms in holographic renormalization and can be regarded as the effective action of ``quantum gravity" on the branes \cite{Chen:2020uac}. As a test of our method, we successfully reproduce the HEE for curvature-squared gravity. Then, we apply this method to derive the HEE for cubic BWHD gravity, which is the main result of this paper. We verify our cubic entropy formula (\ref{sect 3:covariant entropy for L3}) agrees with that of \cite{Miao:2015iba} for $d=6$, which is the universal term of entanglement entropy for 6d CFT dual to Einstein gravity. Our cubic entropy formula has additional terms for $d\ne 6$ and applies to general dimensions. Although the RHEE method works for only sufficiently large $d$, we can make an analytical continuation of the entropy formula to general dimensions. A strong support for the validity of analytical continuation is that the squared entropy formula (\ref{sect 2: get SL2}) agrees with that of \cite{Dong:2013qoa, Camps:2013zua} for general $d$. Besides, we verify our cubic entropy formula (\ref{sect 3:covariant entropy for L3}) agrees with Wald entropy, HEE without/with extrinsic curvatures, and HEE under field redefinition in general dimensions. The main advantage of our approach is that it is independent of the complicated splitting problem. The disadvantage is that it can only deal with the BWHD gravity. Finally, we derive the HEE for quartic BWHD gravity in flat space with constant extrinsic curvatures. It is interesting to apply our method to calculate the HEE for higher-order BWHD terms and those concerning Gauss-Bonnet/Lovelock gravity. It is also interesting to take our entropy formula to study the quantum effect of gravity and the black hole information paradox. We leave them to future work.

\section*{Acknowledgements}
R. X. Miao acknowledges the supports from the National Natural Science Foundation of China (No. 12275366). 

\appendix

\end{document}